\documentstyle[12pt,aaspp4,flushrt]{article}

\lefthead{Klessen, R.~S. and Burkert, A.}
\righthead{The Formation of Stellar Clusters}

\begin{document}

\title{The Formation of Stellar Clusters:  Gaussian Cloud Conditions I}
    \author{Ralf S. Klessen$^{1,2}$ and Andreas Burkert$^{1}$\\ 
{$^{1}$\small  Max-Planck-Institut f\"{u}r Astronomie, K\"onigstuhl 17, 69117
    Heidelberg, Germany}\\
{$^{2}$\small  Sterrewacht Leiden, Postbus 9513, 2300-RA Leiden,
    The Netherlands} }

\begin{center}
{\em (accepted for publication in {\em ApJSS})}\hfill
\end{center}

\begin{abstract}
The isothermal dynamical evolution of a clumpy molecular cloud region
and its fragmentation into a protostellar cluster is investigated
numerically. The initial density distributions are generated from
different realizations of a Gaussian random field with power spectrum
$P(k) \propto k^{-2}$.  During the  evolution of the system, the
one-point probability distribution functions (pdf) of the gas density and of
the line-of-sight velocity centroids develop considerable distortions
away from the initial Gaussian behavior. The density pdf can be best
described by power-law distributions, whereas the velocity pdf
exhibits extended tails. As a result of the interplay
between gas pressure and gravitational forces, a quasi-equilibrium
clump mass spectrum emerges with a power-law distribution $dN/dM
\propto M^{-1.5}$. Being part of a complex network of filaments,
individual clumps are elongated, centrally condensed objects with 2:1
to 4:1 axis ratios with outer $r^{-2}$ density distributions.

Dense, Jeans-unstable gas clumps collapse and form protostellar cores
which evolve through competitive accretion and $N$-body interactions
with other cores.  In contrast to the clumps, the core mass spectrum
is best described by a log-normal distribution which peaks
approximately at the average Jeans mass of the system.  Scaled to
physical conditions typical for star-forming molecular clouds, the
mass function is in good agreement with the IMF of multiple stellar
systems.

The final dynamical state of the newly formed stellar cluster closely
resembles observed young stellar clusters. It has a core/halo
structure which is typical for collision dominated $N$-body systems. The
2-point correlation function of the spatial stellar distribution can
be described by two power-laws with a break in the slope at the
transition point from the binary to the large-scale clustering
regime. The protostellar cluster is marginally bound and would be
easily disrupted, if the conversion of cores into stars is
inefficient.
\end{abstract}


\renewcommand{\thefootnote}{\fnsymbol{footnote}}

\section{Introduction}

The formation of stellar clusters is yet an unsolved problem of
theoretical astrophysics.  Clusters form in giant molecular clouds
complexes on time scales of order $10^6$ yrs (Carpenter et al. 1997,
Hillenbrand, 1997). This time scale is similar to the dynamical time
scale of the clouds, it is however one order of magnitude shorter than
the inferred cloud lifetime (Blitz \& Shu 1980).  The mechanisms that
stabilize molecular clouds for such a long time are not well
understood.  Turbulence has been proposed to stir clouds, supporting
them against gravitational collapse (Arons \& Max 1975). Indeed,
molecular emission lines tend to be an order of magnitude broader than
the thermal linewidth (Blitz 1993), indicating supersonic velocities.
Magnetic fields have been detected by direct measurements of field
strength in some star forming regions (Myers \& Goodman 1988, Crutcher 
et al. 1993 \& 1999, Crutcher 1999, see however Verschuur 1995a,b) and may
affect the cloud dynamics and stability.

Recent simulations suggest that freely decaying turbulence dissipates its kinetic energy
on time scales
shorter than a dynamical time scale (Gammie \& Ostriker 1996, Mac Low et al. 1998, Stone, Ostriker \& Gammie 1998), regardless of the adopted equation of state or the presence of magnetic fields. 
In order to stabilize clouds for a longer time, energy must be supplied, either
by internal processes like stellar winds and outflows (Franco \& Cox 1983, McKee 1989) or externally,
e.g. by the shear motions induced through Galactic differential rotation (Fleck 1981).
The spread in stellar ages for a given cluster is however similar to the
dissipation time scale indicating that clusters form in regions where
turbulence could decay leading to gravitational collapse. How turbulence is driven
and why at some point in the evolution of the cloud the driving mechanism fails 
remains an outstanding issue.

Previous numerical models of isolated gaseous spheres have shown that
stellar clusters could form as a result of gravitational collapse and
fragmentation (see e.g. Larson 1978, Keto, Lattanzio \& Monaghan
1991).  These models were however strongly constrained by numerical
resolution.  More recently, Whitworth et al (1995), Turner et
al. (1995) and Bhattal et al. (1998) investigated in detail the
fragmentation of shocked interfaces of colliding molecular clumps into
small stellar systems. The effects of gas accretion on the evolution
of a newly formed, young stellar clusters has been investigated by
Bonnell et al. (1997).  They found that gas accretion is highly
non-uniform with a few stars accreting significantly more than the
rest and concluded that competitive accretion processes play an
important role in shaping the initial stellar mass function.  Finally,
the individual collapse of perturbed protostellar cores and their
break-up into binaries and multiple systems has recently been studied
with high resolution (see e.g. Burkert \& Bodenheimer 1993, 1996;
Burkert, Bate \& Bodenheimer 1997; Boss 1997, Truelove et al. 1997,
1998, Bate 1998).

In this paper we extend previous studies of the collapse of isolated
gas clumps to the molecular cloud regime and study the formation of
stellar clusters localized within globally stable molecular
clouds. Our simulations combine self-consistently previous studies of
the dynamics of turbulent molecular clouds with studies of their
collapse and fragmentation stage till the formation of a stellar
cluster including its competitive accretion phase.  In a previous
short Letter (Klessen et al. 1998) we discussed a high-resolution
simulation which lead to a stellar cluster with a log-normal mass
distribution in excellent agreement with observations of multiple
stellar systems.  The current paper will explore the formation and early
evolution of a star cluster in greater detail, starting from different
initial realizations of clumpy molecular clouds with initial
power-spectrum $P(k) \propto k^{-2}$. Section 2 presents the numerical
method. Model properties and initial conditions are presented in
section \ref{sec:scaling}. The global evolution of the models is
discussed in section \ref{sec:time-evolution}.  Section
\ref{sec:discussion} analyzes the dynamics and structure of the gas
cloud and the properties the newly formed stellar system and compares
it with observations. Our results are summarized in section \ref{sec:summary}.

\section{The numerical model}
\label{sec:numerics}
\subsection{SPH in combination with GRAPE}
SPH ({\em smoothed particle hydrodynamics}) is a particle-based scheme
to solve the equations of hydrodynamics.  As the fluid is represented
by an ensemble of particles, the technique can be regarded as an
extension of the pure gravitational $N$-body system.  Besides being
characterized by its mass $m_i$, velocity ${\vec v}_i$ and location
${\vec r}_i$, each particle $i$ is associated with a density $\rho_i$,
an internal energy $\epsilon_i$ (equivalent to a temperature $T_i$),
and a pressure $p_i$. The time evolution of the fluid is represented
by the dynamical evolution of the SPH particles. Their behavior is
governed by the equation of motion supplemented by additional
equations to modify the hydrodynamic properties.  Thermodynamic
observables are obtained by averaging over an appropriate subset of
the SPH particles. Excellent overviews over the method, its numerical
implementation, and some of its applications have been written by Benz
(1990) and Monaghan (1992).  We use SPH because it is intrinsically
Lagrangian: As opposed to mesh-based methods, it does not require a
fixed grid to represent fluid properties and calculate spatial
derivatives (see e.g.~Hockney \& Eastwood 1988). The fluid particles
are free to move and -- in analogy -- constitute their own grid. The
method is therefore able to resolve very high density contrasts
because the particle concentration increases where needed.

Our code is based on a version originally developed by Benz (1990).
It uses a standard description of a von~Neumann-type artificial
viscosity (Monaghan \& Gingold 1983) with adopted parameters
$\alpha_{\rm v} = 1$ and $\beta_{\rm v} = 2$ for the linear and
quadratic terms. The system is integrated in time using a second-order
Runge-Kutta-Fehlberg scheme, allowing individual time steps for each
particle. In any self-gravitating fluid, regions with masses exceeding
the Jeans limit become unstable and collapse. In the current code,
once a highly-condensed object forms in the center of a collapsing gas
clump and has passed a certain density threshold, the dense core is
substituted by a `sink' particle (Bate, Bonnell \& Price 1995). This
particle has a fixed accretion radius, which is of order of the Jeans
length at the density at which the sink particle is created. It
inherits the combined masses, linear and `spin' angular momenta of the
particles it replaces. It has the ability to accrete further SPH
particles from its infalling gaseous surrounding. Again, the mass of
the accreted particles, their linear and angular momenta will be added
to the sink particle in order to guarantee mass and momentum
conservation.  The accreted particles are then removed from the
calculation.  By adequately replacing high-density cores by sinks and
keeping track of their further evolution in a consistent way the code
time stepping is prevented from becoming prohibitively small and we
are able to follow the dynamical evolution of the system over many
free-fall times. However, this procedure implies that information
about the evolution of gas inside the sink particle is lost. In our
case a sink particle corresponds to a gravitationally collapsing
protostellar core.  For a detailed description of the physical
processes inside a protostellar core, a new simulation just
concentrating on this single object would be required with the
appropriate initial and boundary conditions taken from the
larger-scale simulation (Burkert, Klessen \& Bodenheimer 1998).

To achieve high computational speed, we use SPH in combination with
the special-purpose hardware device GRAPE (Sugimoto et al.~1990,
Ebisuzaki et al.~1993).  This device calculates the forces and the
potential in the gravitational $N$-body problem by direct summation on
a specifically designed chip with very high efficiency. This allows
calculations at supercomputer level on a normal
workstation. Additionally, GRAPE returns the list of nearest neighbors
for each particle. This feature makes it attractive for use in
smoothed particle hydrodynamics (Umemura et al.~1993, Steinmetz
1996). For particles near the surface of the integration volume,
`ghost' particles are created to correctly extend the neighbor search
beyond the borders of the computational volume; no forces are computed
for these particles.

Since we wish to describe the dynamical evolution of a gravitationally
unstable region in the interior of a considerably larger globally
stable molecular cloud, we adopt periodic boundary conditions to
prevent overall collapse.  As {\sc Grape} cannot treat periodic
particle distributions directly due to its restricted force we have to
introduce the Ewald (1921) method to prevent global collapse.  The
basic idea is to compute a periodic correction force for each particle
on the host computer, applying a particle-mesh like scheme: We first
compute the forces in the isolated system using direct summation on
{\sc Grape}, then we assign the particle distribution to a mesh and
compute the periodic correction force for each grid point, by
convolution with the adequate Green's function in Fourier
space. Finally, we add this correction to each particle in the
simulation. The corrective Green's function can be constructed as the
offset between the periodic solution (calculated via the Ewald
approximation) and the isolated solution on the grid. This method has
proven to be numerically stable and inexpensive in terms of the
computational effort (Klessen 1997).

\subsection{Some cautionary remarks on the limitations of SPH}
\label{subsec:resolution-limit}
To make full use of the Lagrangian nature and resolving power of SPH
the smoothing volume over which hydrodynamic quantities are averaged
in the code is freely adjustable in space and time such that the
number of neighbors considered is always kept in the range 30 and 70,
with the optimum value being 50.  This sets a natural limit to the
spatial resolution of the code. It is furthermore constrained by the
Courant-Friedrichs-Lewy (1928) criterion.  It demands that the minimum
time stepping in the SPH code is always less than the time required
for a sound wave to cross the minimum smoothing volume. In order to
prevent the time stepping required to resolve very high density peaks
to become prohibitively small, one has to introduce a minimum
smoothing length which defines the smallest resolvable length scale.
In our simulations of self-gravitating gas the spatial resolution is
subject to an additional constraint. In order to correctly treat the
dynamical evolution of high-density peaks, the mass contained within
the smoothing volumes of two interacting particles must be less than
the local Jeans mass. Otherwise, the stability of the clump against
gravitational collapse depends on the detailed implementation of the
gravitational force law and on the kernel function used for the
simulation, rather than on physical processes.  The minimum Jeans mass
that is reached during the calculation must always be greater than
approximately twice the mass of particles in the SPH kernels (Bate \&
Burkert 1997). If one bears this caveat in mind, the SPH method
calculates the time evolution of gaseous systems very reliably and
accurately. Additionally it offers spatial and dynamical flexibility
that has yet to be achieved by grid-based methods.

\section{The model}
\label{sec:scaling}
The numerical models discussed in this paper describe
self-gravitating, isothermal gas. This is justified by the typical
densities observed in molecular clouds and determines the equation of
state as well as the physical processes considered in the model. The
dynamical evolution of the gas is scale-free and depends only on the
ratio of the internal to gravitational energy. We start with an
initially clumpy cloud region with density fluctuations following a
Gaussian random distribution. The velocities are computed
self-consistently from the Poisson equation. Assuming that stellar
clusters form through the gravitational collapse of clumpy cloud
regions that have lost their turbulent support, we study the detailed
behavior of this process and the properties of the newly formed
cluster.

\subsection{Scaling properties of isothermal, self-gravitating gas}
\label{subsec:scale-free}
For isothermal gas, the energy density is a function of temperature
only and the equation of state reduces to $p = c^2_{\rm s}\;\!  \rho$,
with $c_{\rm s}$ being the thermal sound speed. This approximation is
valid for physical regimes where the cooling time scales are much less
than the dynamical ones. In molecular clouds, this is the case for gas
densities $1 \lesssim n({\rm H}_2) \lesssim 10^{10}\,$cm$^{-3}$, where
the gas is optically thin for the dominant cooling processes and
energy is radiated away very efficiently (e.g.~\cite{toh82}).  Average
densities in star forming regions typical are in the range $10\,{\rm
cm}^{-3} < n({\rm H}_2) < 10^5\,{\rm cm}^{-3}$ which is the density
range of our models.  Even when resolving a density contrast of
$10^4$, the isothermal equation of state is  appropriate
throughout the entire simulation.

The self-gravitating, isothermal model studies the interplay between
gravity and gas pressure, it is thus scale free. Besides the dependence on the
initial density and velocity distribution, the dynamical evolution of
the system depends only on one additional free parameter, the ratio between
the internal energy $\epsilon_{\rm int}$ and the potential energy
$\epsilon_{\rm pot}$. This ratio can be interpreted as a dimensionless
{\em temperature},
\begin{equation}
\alpha \equiv \epsilon_{\rm int}/|\epsilon_{\rm pot}|\:.
\end{equation}
Line widths in molecular clouds are super-thermal, implying the
presence of supersonic turbulent motions (e.g.~Blitz 1993). In case of
isotropic turbulence, these non-thermal (turbulent) contributions can
be accounted for by introducing an {\em effective} energy
$\epsilon_{\rm int} = \epsilon_{\rm therm} + \epsilon_{\rm turb} =
\gamma\cdot(c^2_{\rm s} + \sigma^2_{\rm turb})/2$ adding up the
thermal and non-thermal contributions to the kinetic energy. This is
equivalent to defining an {\em effective} temperature $\alpha_{\rm
eff}$.  The turbulent velocity dispersion is denoted by $\sigma_{\rm
turb}$ and the factor $\gamma$ depends on the degree of freedom. In
case of anisotropic turbulent motions, the system has (locally)
preferred axes and the concept of one single effective temperature is
not valid.

\subsection{Normalization and relation to observed star-formating regions}
\label{subsec:conversion}

In analogy to the temperature, we adopt dimensionless and normalized
units for all physical quantities.  All masses are scaled relative to
the total mass of the simulated molecular cloud region, and the length
scale is its size $L$.  The numerically simulated area is then a cube
$[-L,+L]^3=[-1,+1]^3$ with periodic boundary conditions. The density
of the homogeneous cube is $\rho = 1/8$.  In order to trigger
gravitational collapse and star formation we consider systems that are
highly unstable against gravitational collapse and set $\alpha =
0.01$. The volume will then contain $N_J=222$ Jeans masses.  We assume
the gravitational constant $G\equiv1$ and the gas constant ${\cal R} =
1/\gamma$ with $\gamma=3/2$ for an ideal gas having three degrees of
freedom. In these units the sound speed is $c_{\rm s} = ({\cal R}
\alpha)^{1/2}$ and the Jeans mass follows as
\begin{equation}
M_{\rm J} = 1.6\cdot\rho^{-1/2}\;\alpha^{3/2}\;.
\end{equation}

In order to scale to physical units let us consider a dark cloud like
Taurus with $n({\rm H}_2) \approx 10^2\,{\rm cm}^{-3}$ and
$T\approx10\,$K.  Assuming a mean molecular weight of $\mu = 2.36$,
the units of mass and length correspond to $M = 6\;\!300\,$M$_{\odot}$
and $L = 5.2\,$pc, respectively. The time unit is equivalent to
$t=2.2\times10^6\,$years and the average Jeans mass transforms to $
M_{\rm J} = 28\,{\rm M}_{\odot}$. Applied to a dense, massively
star-forming cloud with $n({\rm H}_2) \approx 10^5\,{\rm cm}^{-3}$ and
$T\approx10\,$K, similar to the BN region in Orion, the simulated cube
translates into a mass of $M = 200\,$M$_{\odot}$ and a size scale $L =
0.16\,$pc. The time unit now converts to $t=7.0\times10^4\,$years and
the mean Jeans mass for the homogeneous distribution is $M_{\rm J} =
0.9\,{\rm M}_{\odot}$.

Molecular clouds are stabilized against global collapse by the
presence of supersonic turbulence as indicated by the large observed
line width (e.g.~Blitz 1993).  However, molecular clouds do form
stars.  Star formation occurs in their interior in regions which {\em
locally} lose turbulent support and which begin to contract to form
stars or clusters of stars.  The time scale for freely decaying
turbulence is of the order of the free-fall time or shorter (e.g.~Mac
Low et al.~1998). As the processes that lead to the local loss of
turbulent support are not well understood, we do not attempt to model
that stages of the evolution and start our simulations when turbulence
has already decayed.  This is equivalent to the assumption of
instantaneous loss of support.  The sizes of local collapse regions
are typically much smaller than the overall extent of the cloud. To a
good approximation we therefore neglect the influence of cloud
boundaries and place the considered volume inside an infinite
cloud. Hence, to describe a gravitationally unstable volume inside an
overall stable cloud of much larger extent we adopt periodic boundary
conditions. This is a commonly used scheme when describing turbulent
cloud dynamics (e.g. Mac Low et al. 1998, Stone et al. 1998, Klessen
2000).  

With the adopted values of $\alpha = 0.01$ collapse progresses quite
rapidly and the influence of the boundary conditions on the overall
dynamics is weak. The situation changes when the scaling parameter
$\alpha$ is increased. The total number of Jeans masses contained in
the computed cube then decreases, which in our isothermal models is
equivalent to `zooming' in onto smaller spatial volumes.  Hence, 
the adopted boundary conditions become more important and may begin to
influence the collapse behavior of individual protostellar cores.  As
cores are never isolated, but instead interact with each other or
accrete gas from their environment, in the case of $\alpha
{\:\lower0.6ex\hbox{$\stackrel{\textstyle >}{\sim}$}\:} 0.1$
boundaries allowing for mass inflow would become more appropriate.  Indeed,
this study may be used for finding the optimum boundary conditions for
high-resolution simulations which focus on the evolution of individual
protostellar cores. We defer a detail discussion of this issue to
paper II where we address the influence of different initial and
boundary conditions. Decreasing $\alpha$ on the other hand means
increasing the corresponding physical extend of the computed
volume. The simulation treats a larger fraction of molecular cloud
material and results in a larger number of protostellar cores that
form during the dynamical evolution and fragmentation of the gas.
This, however, also increases the computational demands (if a fixed mass
and spatial resolution for individual cores is retained). The adopted
value $\alpha = 0.01$ in the present paper corresponds to the maximum
number of Jeans masses which we can handle with sufficient numerical
resolution.

\subsection{Initial conditions -- Gaussian random fields}
\label{subsec:initial}
The dynamical evolution of gas in a molecular cloud region depends on its
initial density and velocity distribution. In a self-consistent model,
both are related via Poisson's and the hydrodynamic equations.  We
adopt random Gaussian fluctuations for the density as starting
condition for the SPH simulations. We use this approach, because these
distributions have well determined statistical properties and can
easily be generated by the Zel'dovich approach (see appendix
\ref{app:zel}). Furthermore, their properties resemble the end stages
of decaying turbulence and are suited to mimic observed features of
molecular clouds once advanced into the non-linear regime (see
e.g.~Stutzki \& G{\"u}sten 1990, who deploy a Gaussian decomposition
technique to describe the clumpy structure of molecular clouds).

Gaussian random fields $\rho(\vec{r})$ are completely characterized by
their first two moments, the mean value $\rho_0 \equiv \langle
\rho(\vec{r})\rangle_{\scriptsize \vec{r}}$ and the 2-point
correlation function $\xi(\vec{r}) \equiv \langle
\rho(\vec{r}')\rho^*(\vec{r} +\vec{r}')\rangle_{\scriptsize
{\vec{r}'}}$, which is equivalent to the power spectrum $P(\vec{k})$
in Fourier space. All higher moments can be expressed in terms of
$\xi(\vec{r})$.  For an isotropic fluctuation spectrum, $P(k) =
P(|\vec{k}|)$, the 2-point correlation degrades to a function of
the distance between two points, $\xi(r)$.  By defining a
normalization $\rho_0$ and the power spectrum $P(k)$ in Fourier space,
all statistical properties of the field $\rho(\vec{r})$ are
determined.

The function $P(k)$ identifies the contribution of waves with wave
number $k$ to the statistical fluctuation spectrum. In Gaussian random
fields, the phases are arbitrarily chosen from a {\em uniform}
distribution in the interval $[0,2\pi[$, and the amplitudes for each
mode $k$ are randomly drawn from a {\em Gaussian} distribution with
width $P(k)$ centered on zero. Since waves are generated from random
processes, only the properties of an {\em ensemble} of fluctuation
fields are determined in a statistical sense. Individual realizations
(from different sets of random numbers) may deviate considerably from
this mean value, especially at small wave numbers $k$, i.e.~at long
wave lengths, where only a few modes $(k_x,k_y,k_z)$ contribute to the
wave number $k=|\vec{k}|$. This means that different density fields
generated from the same power spectrum $P(k)$ may {\em look} quite
differently, especially on large scales (see
Fig.~\ref{fig:nine-models}), despite having identical statistical
properties. This variance effect can be reduced by considering and
averaging over a large enough ensemble of realizations.  We generate
the initial density fluctuation fields by applying the Zel'dovich
(1970) approach. The method and its applicability to gaseous systems
is discussed in detail in the appendix \ref{app:zel}.

In the current paper, we concentrate our analysis on a simple
power-law functional form for the fluctuation spectrum, $P(k) \propto
k^{\nu}$, with ${\nu} = -2$.  As ${\nu}$ is a negative number, most
power is in large-scale modes. This choice gives a good representation
of the observed patchiness and inhomogeneity of molecular clouds. The
dependence of the dynamical evolution of the model cloud on variations
of the initial density distribution with different exponents ${\nu}$
of the power law will be discussed in  paper II.

\subsection{Model realization}
\label{subsec:model-properties}
We present results from a
detailed analysis of nine different initial realizations of $P(k) \propto k^{-2}$
and use different particle numbers to address the
issue of numerical resolution: six models with $50\!\;000$ SPH
particles, two with $200\!\;000$ and one with $500\!\;000$ particles,
respectively. All models are generated using the Zel'dovich (1970)
approximation (see appendix \ref{app:zel}) and their properties are summarized in
Tab.~\ref{tab:models-N=2}.  The projection into the $xy$-plane of the
initial particle distribution in each model at the start of the
dynamical evolution with SPH is presented in
Fig.~\ref{fig:nine-models}.

All but one model assume a random uniform particle distribution before
applying the Zel'dovich shift. Placing particles randomly into a given
volume produces an overall distribution that is homogeneous on large
scales, but is subject to statistical fluctuations on small scales
which introduce white noise into the correlation function. The scale
at which this effect becomes important is of the order of the mean
inter-particle distance. These undesired small-scale fluctuations may
influence the fragmentation behavior of the gas. However, with the
adopted temperature, the mass of small-scale regions of enhanced
density is much less than the local Jeans mass. Therefore, these
fluctuations are quickly damped and dynamically unimportant.  As
alternative to the random distribution, in model $\cal F$ particles
are placed on a regular grid before applying the Zel'dovich
approximation. This distribution is force-free and has exactly uniform
density. On the other hand, the grid has preferred axes which
introduce anisotropies in the neighbor list of each particle. This
fact may again influence the small-scale behavior. On larger scales,
the system is increasingly isotropic and this effect becomes
negligible. In the hydrodynamic evolution phase after the Zel'dovich
shift traces of the grid are quickly erased. Comparing the dynamical
evolution, the model with the particles placed on a grid is
statistically indistinguishable from the models generated from a
random distribution, which we use as standard.

As discussed in appendix~\ref{app:zel}, the Zel'dovich shift interval
$\delta t$ determines the density contrast in the particle
distribution. Typically, larger $\delta t$ leads to higher initial
peak densities. On the other hand, smaller $\delta t$ implies that a
larger fraction of the total time evolution of the gas has to be
computed with the SPH method and pressure forces, which are not
included in the Zel'dovich approximation, have more time to act on the
gas.  In general, $\delta t$ should be chosen small enough so that the
subsequent evolution is not dependent on the choice of $\delta t$. In
our models this is indeed the case for $\delta t \lesssim 2.0$ as is
shown in appendix \ref{subsec:validity}. To address this issue further and
to examine how a variation of the Zel'dovich shift interval influences
the properties of the protostellar cluster that forms during the
dynamical evolution and collapse of the gas, three models are
generated with smaller shift intervals $\delta t$: model $\cal E$ with
$\delta t = 1.0$, and model $\cal H$ together with the high-resolution
model $\cal I$ having $\delta t = 1.5$. Within the statistical
variance between different models the results are not dependent on the
initial shift interval.

\section{Global time evolution}
\label{sec:time-evolution}
Due to the small value of $\alpha$ the gravitational energy outweighs
the internal energy by far and the system is highly unstable to
gravitational collapse and fragmentation.  As a result, typically
about sixty collapsed cores form during the dynamical evolution. A
complete time evolution is illustrated in
Fig.~\ref{fig:3D-cube-A}. Representative for all nine models it shows
snapshots of the model $\cal A$ at twelve different stages of its
dynamical evolution.  Note that the cube has to be seen periodically
replicated in all directions. With the start of the SPH simulation,
pressure forces begin to act on the gas and smear out fluctuations
which are below the local Jeans limit. On the other hand, large modes
are unstable against gravitational collapse and start to contract. At
$t\approx 0.4$ the first highly-condensed cores form in the centers of
the most massive and densest Jeans-unstable gas clumps and are
replaced by sink particles (see Sec.~\ref{sec:numerics}).  Soon,
clumps of smaller initial mass and density follow. The density
threshold for the formation of sink particles is chosen to be
$\rho_{\rm c} = 5000$ and the diameter of the sink particles is 1/100
of the linear size of the simulated cube.  It is visible in
Fig.~\ref{fig:3D-cube-A} that the system evolves into a network of
intersecting sheets and filaments. The gas density is highest along
filaments and at their intersections. These are the locations where
dense cores form predominantly and soon a hierarchically-structured
cluster of accreting protostellar cores, represented by sink particles
is built up. Whereas the overall dynamical evolution of the system is
initially dominated by hydrodynamic effects (all the mass is in the
gas phase), the later evolution is increasingly determined by the
gravitational $N$-body interaction between protostellar cores and
between cores and their clumpy gaseous environment because more and
more gas is accreted onto dense cores. The final result is a bound
dense cluster of protostars.  After a few cluster crossing times, core
motions have been randomized by close encounters and the protostars
have lost knowledge of their initial conditions.

At this stage it is necessary to draw attention to one caveat: The gas
in the models is treated isothermal. This implies that there exists no
feedback mechanism which may prevent the complete accretion of all the
gas into cores and at late phases the global core-formation
efficiency\footnote{We use the word core-formation efficiency to
distinguish from the commonly quoted star-formation efficiency. We
identify the sink particles with unresolved, collapsing protostellar
cores.  Stars and cores are connected via the ability and
effectiveness of {\em individual} cores to form stars.}, defined by
the percentage of gas that is converted into dense cores, approaches
unity. This appears unphysical and is not observed in star-forming
regions.  At some stage during the protostellar evolution phase,
feedback processes from nascending stars inside protostellar cores or
from already existing massive stars in their vicinity will become
relevant and terminate the accretion and collapse phase. These effects
cannot be treated in an isothermal model. However, the early
(isothermal) phases of the collapse of molecular cloud regions and the
formation of a stellar cluster are well described. The time at which
these assumptions become less appropriate is difficult to estimate. We are
therefore hesitant to interpret the models beyond the phase at which
more than $\sim60\;\!$\% of the gas has been accreted onto
protostellar cores. 

In Fig.~\ref{fig:3D-cube-I}, four stages of the evolution of the
high-resolution model $\cal I$ are
presented. Figure~\ref{fig:3D-cube-I} plots the system
initially, at time $t=1.4$, when 10\% of the gas is condensed into
protostellar cores and t=2.0 and t=2.8 when 30\% and 60\%
of the total mass has been accreted, respectively. 
Instead of placing individual SPH particles,
Fig.~\ref{fig:3D-cube-I} plots the distribution of the gas
density. The density field is scaled logarithmically with darker areas
denoting higher densities. Hence, dark dots identify the location of
dense collapsed cores.  When comparing the time scales for core
formation and subsequent accretion with the previous model, the
different Zel'dovich shift intervals have to be taken into account. In
the high-resolution calculation $\cal I$ it is chosen to be $\delta t=
1.5$ instead of $\delta t=2.0$ for model $\cal A$. Furthermore, the
adopted accretion radius of sink particles in the code was reduced by
a factor of two, it is now $1/200$ of the linear size of the simulation
box. This delays the formation and accretion onto sink particles by
$\Delta t\approx 0.3$. Altogether, the state of the system where the mass
fraction accumulated in collapsed cores exceeds a certain percentage
is reached somewhat later by $\Delta t \approx 0.8$ in simulation
$\cal I$. Besides
this delay, the dynamical evolution of both systems is very
similar, leading to the formation of a cluster of protostellar 
cores which grow in
mass by accretion from its surrounding gas reservoir.
We conclude that our basic results are not resolution dependent.

\section{Discussion}
\label{sec:discussion}
Supplementing the general outline of the dynamical evolution in the
last section, we now present a detailed analysis of the large-scale
collapse and fragmentation behavior of the system. First, we discuss
the one-point probability density functions of the density and of the
distribution of line-of-sight velocity centroids. Then we describe the
clumping properties of the gas at various stages of its dynamical
evolution, the kinematic and spatial properties of the protostellar
cluster that forms as the system advances in time and the mass
spectrum of protostellar cores. Furthermore, we discuss the
boundedness of the cluster and the rotational properties of
protostellar cores. Finally, we speculate about the implications of
our results for understanding the initial stellar mass function (IMF)
by comparing the numerically calculated core mass spectrum with the
IMF for multiple stellar systems.

\subsection{Evolution of the one-point probability density functions for
density and line-of-sight velocity centroids}
\label{subsec:pdfs}
There has been a series of recent attempts to use one-point
probability density functions (pdf) as statistical descriptors of
properties of the interstellar medium, in particular as diagnostic to
distinguish between different physical processes influencing its
dynamics and time evolution. For a discussion of the one-point
probability density function of the density field and its relevance
for the star formation process see e.g.~Scalo et al.~(1998a) and
references therein. Detailed analyses of the pdf of line centroid
velocity fluctuations are presented by Lis et al.~(1996, 1998),
Miesch, Scalo \& Bally (1998), and Klessen (2000).  Here we derive
both sets of functions for our model of larger-scale cloud collapse.

\subsubsection{The density pdf}
We define the pdf of gas densities, $f(\rho)$, such that
$f(\rho)d\rho$ measures the fraction of the total mass in the system
that falls into the density range $[\rho, \rho+d\rho]$. The function
$f(\rho)$ has of inverse density. Basically it is the histogram of the
density summed over all SPH particles in the simulation. This defines
$f(\rho)$ as {\em mass} weighted average. Note that some authors
define the function by measuring the fractional {\em volume} occupied
by gas within the density range $[\rho,\rho+d\rho]$. The later
definition is convenient when analyzing numerical simulations
performed with a grid-based method.  SPH on the other hand is a
particle based method, therefore it is more appropriate to obtain
statistical measures by summing over the contributions of individual
particles and hence obtain mass weighted averages.

Pure hydrodynamic simulations of turbulent isothermal gas {\em
without} self-gravity or additional physical processes result in
density pdf's of log-normal functional shape, i.e.~the logarithm of
the density follows a normal (Gaussian) distribution. However, there
is increasing evidence that the inclusion of additional physical effects
(like self-gravity or turbulent driving mechanisms) will lead to
density pdf's which significantly deviate from being log-normal (Scalo
et al.~1998, Passot \& V{\'a}zquez-Semadeni 1998, \cite{nordlund98},
Klessen 2000, Mac~Low \& Ossenkopf 2000).

This is also found in our simulations. Using data from the
high-resolution simulation $\cal I$, Fig.~\ref{fig:rho-pdfs} clearly
shows that the evolution of Jeans unstable isothermal self-gravitating
gas leads to {\em non}-log-normal density pdf's
which can be described as power laws during most evolutionary
stages. The initial state of the SPH evolution ($t=0$,
Fig.~\ref{fig:rho-pdfs}a) still exhibits a well established log-normal
density pdf. This is expected after applying the Zel'dovich shift to
transfer a Gaussian fluctuation spectrum -- which has {\em per
definitionem} a normal pdf in the linear regime -- into the non-linear
regime (\cite{kofman94}). Later on self-gravity and gas pressure begin
to shape the mass distribution into a network of intersecting sheets
and filaments and ultimately into a cluster of collapsed dense cores
(see Fig.~\ref{fig:3D-cube-A} and \ref{fig:3D-cube-I}). More and more
gas converges into regions of high density and consequently the
density pdf develops a high-density `tail' of a power-law functional
form. The power-law distribution is best established at that stage of
the evolution, when the first and most massive Jeans unstable gas
clumps have collapsed to high enough central densities that their
innermost regions can be identified as compact cores (and subsequently
are substituted by sink particles in our code).  This happens at
$t \approx 1.2$ and leads to a slope of the distribution function of
$d\log_{10}N/d\log_{10}\rho = -0.8$ as can be seen in
Fig.~\ref{fig:rho-pdfs}b. Densities of $\log_{10} \rho \approx 4$ in
this figure arise from the contributions of the collapsed cores. For
those, the values of $\log_{10} \rho$ are lower limits since they are
computed by dividing the masses of the cores by their (fixed) volume;
isothermal gas would continue to collapse to ever increasing central
densities (until the gas becomes optically thick and begins to heat
up). Subsequently, with the number and masses of collapsed cores
increasing, the density distribution is best
described by multiple power laws. A steep
part describes collapsing gas that forms a new core or that is
accreted onto dense cores, and a shallower part that describes matter
that does not (yet) participate in collapse and fills the 
low-density regions between filaments and
cores (see Fig.'s \ref{fig:rho-pdfs}c -- e). In the final stage of the
evolution (Fig.~\ref{fig:rho-pdfs}f) almost all of the mass is contained in a
cluster of dense protostellar cores (with densities of $\log_{10}
\rho \approx 4$) immersed into gas of low density exhibiting an almost
flat density pdf.

In general, the spread and shape of pdf's can be characterized by using their
statistical moments. For each of the six epochs of simulation $\cal I$
that are shown in Fig.~\ref{fig:rho-pdfs} we compute
the first four moments which are defined in the following way:
\begin{eqnarray}
{\rm mean} & = \,\mu\, = & \frac{1}{N}\sum_{i=1}^N \rho_i\;,\label{eqn:mom1}\\
{\rm standard \;\;deviation} & =\, \sigma \,= & \sqrt{\frac{1}{N}\sum_{i=1}^N (\rho_i-\mu)^2}\;,\label{eqn:mom2}\\
{\rm skewness} & =\, \theta \,= & \frac{1}{N}\sum_{i=1}^N \frac{(\rho_i-\mu)^3}{\sigma^3}\;,\label{eqn:mom3}\\
{\rm kurtosis} & =\, \kappa \,= & \frac{1}{N}\sum_{i=1}^N \frac{(\rho_i-\mu)^4}{\sigma^4}\;.\label{eqn:mom4}
\end{eqnarray}
The summations extend over all $N=500\,000$ SPH particles in the
system. Mean $\mu$ and standard deviation $\sigma$ quantify the
location and the width of the pdf and are given in density units.  The
third and fourth moments are dimensionless quantities characterizing
the shape of the distribution. The skewness $\theta$ describes the
degree of asymmetry of the distribution around its mean. A negative
value indicates an asymmetric tail extending towards the negative
side, a positive value indicates a positive tail in the distribution.
The kurtosis $\kappa$ measures the relative peakedness or flatness of
the distribution. Gaussian distributions are characterized by
$\kappa=3$. Smaller values indicate distributions that have flatter
peaks, and larger ones point towards strong peaks or equivalently the
existence of prominent tails. A pure exponential distribution leads to
$\kappa=6$. The values for model $\cal I$ are listed in
Tab.~\ref{tab:moments}. Since we analyze a mass weighted density
distribution, the first moment $\mu$ is not identical to the global
average density, $\rho = 1/8$. Most of the mass is in gas of larger
densities. Using only the first two moments (i.e.~a Gaussian fit) to
reconstruct the logarithmic density pdf never leads to a good
representation of the data except for the initial configuration
(Fig.~\ref{fig:rho-pdfs}a).

In summary, the filamentary morphology of the gas distribution and the
power-law appearance of the density pdf in our models is clearly a
result of the presence of self-gravity which leads to collapse and
compression and thus to the occurance of high density contrasts. The
hydrodynamic evolution of isothermal gas naturally produces
log-normal density pdf's (see also \cite{passot98}). The inclusion of
self-gravity leads to deviations from this simple analytical form
towards more complex pdf's which during most stages of the dynamical
evolution of the system can be described by power laws at the large
density part of the distribution.

\subsubsection{The line-of-sight velocity centroid pdf}
One way to infer the three-dimensional structure of molecular clouds
from the observed column density distribution is to include the
velocity information along the line-of-sight. Using velocity coherence
as indication for spatial correlation one is able to deconvolve the
cloud into smaller subunits, i.~e.~into clumps on different
scales. This is commonly applied to a variety of molecular clouds
(\cite{lor89}, \cite{stu90}, \cite{willi94}, \cite{oni96},
\cite{kram98}, \cite{heithausen98}) and we will discuss the clump
structure of our models in the next section. However, other
combinations of spatial and velocity information are possible and
useful as well. There are attempts to characterize the structure of
molecular clouds (and the turbulent interstellar medium in general) by
using the pdf's of molecular line centroid velocity fluctuations
(\cite{miesch95}, \cite{lis98}, \cite{miesch98}). The idea is to
determine the centroids of molecular lines for a large number of
different positions across the face of a molecular cloud and from that
compute their distribution function in velocity space. Like in the
case of the density pdf, by doing so direct spatial information is
lost, but the velocity centroid pdf still contains statistical
information about the cloud and may be used to differentiate between
different theoretical models. For instance, it has been shown
experimentally and in numerical simulations that the pdf of the
velocity field for incompressible turbulence is very nearly Gaussian
although non-zero skewness should exist to some degree (e.g.~Vincent
\& Meneguzzi 1991, Cao et al.~1996, Vainshtein 1997, Lamballais et
al.~1997). This situation changes in the regime of highly compressible
supersonic turbulence, the velocity pdf can deviate quite
significantly from the Gaussian form (Lis et al.~1998, Klessen 2000,
Mac~Low \& Ossenkopf 2000).  This behavior is also found in the
simulations of clustered star formation discussed in this paper. Here
it is due to the presence of self-gravity which introduces strong
collapse motions into the velocity field.
 
Again using data from the high resolution model $\cal I$,
Fig.~\ref{fig:vel-pdfs} shows centroid pdf's for the line-of-sight
velocities along the $x$-, $y$-, and $z$-axis at four different stages
of the dynamical evolution of the system, (a) at $t=1.2$ just when the
first dense cores have formed and contain roughly 1\% of the total
mass, (b) at $t=2.0$ when altogether 33 cores have formed and accreted
30\% of the total mass, (c) at $t=2.8$ when 60\% of the available gas
mass has been accreted, and finally (d) at $t=3.9$ when 90\% of the
mass is accounted for by a cluster of 56 cores. We divide the face of
the simulation cube into $50^2$ cells of size $0.04^2$. For each cell,
we compute the line profile by binning the normal (line-of-sight)
velocity components of all gas particles that are projected into that
cell; the width of each velocity bin is $\Delta v = 0.05$ which is
60\% of the sound speed  $c_{\rm s} = 0.08$.  This procedure
corresponds to the formation of optically thin lines in molecular
clouds where all molecules within a certain column through the clouds
contribute to the shape and intensity of the line. We obtain the
centroid of each line and the desired pdf by plotting the histogram
  of all centroid velocities. To reduce the sampling uncertainties, we
repeat the same procedure, but with the location of each cell shifted
by half a cell size. Altogether 5000 lines contribute to the pdf,
each line being determined by the velocities of on average 200 SPH
particles. The left column depicts the pdf for the line-of-sight along
the $x$-axis (i.e.~for particle projections onto the $yz$-plane), and
the middle and right column for the $y$- and $z$-component,
respectively. Note that in the linear-log plots the  Gaussian
distribution has a parabolic shape. The small inlay in the upper
left corner of each histogram gives the {\em overall} line profile
resulting from observing the entire cube, i.e.~including the
line-of-sight velocities of {\em all} gas particles in the
system. Analog to plotting the pdf, the horizontal axis of the inlay gives the
velocity in linear scale (each tick mark denoting a velocity increment
$\delta v = 1.0$) and the vertical axis the logarithm of the number of
particles that contribute to a velocity bin (with the separation of
each tick mark being $\log_{10} N = 1.0$).

Since the thermal sound speed is $c_{\rm s} = 0.08$ all line widths and
the widths of the centroid pdf's are highly supersonic. Furthermore,
none of the pdf's resembles a Gaussian. For comparison we have plotted
for each of the pdf's a Gaussian curve constructed from using the
first two moments of the distribution presented in
Tab.~\ref{tab:moments} (the dashed parabolas in each plot). This curve
never fits any of the pdf's and is typically too small in width. The
distributions show extended tails and can appear quite asymmetric as
indicated by the large values for the third and fourth moments, also
listed in Tab.~\ref{tab:moments}. Furthermore the pdf's show
considerable substructure, for instance notice the double peaked
distribution in Fig.~\ref{fig:vel-pdfs}b ($z$-component) or
Fig.~\ref{fig:vel-pdfs}c ($y$-component). This is a sign of the
streaming motion of gas onto a local center of gravity along filaments
which by chance are aligned with the line-of-sight. Some pdf's show
nicely developed exponential tails, like Fig.~\ref{fig:vel-pdfs}c
($z$-component) or Fig.~\ref{fig:vel-pdfs}d ($y$-component).  As
collapse and star formation progresses in time the width of the pdf's
grows. However, pdf's at equal times but obtained from different
projection angles also exhibit considerable deviations from each
other. Depending on whether one looks along a major axis of
contraction or perpendicular to it the width of the pdf appears larger
or smaller. In addition, massive gas clumps moving through the cloud
will lead to individual features in the velocity pdf. Altogether, the
clumpy structure of molecular clouds causes considerable substructure
in the observed velocity pdf, it will never appear as a smooth function.

One of the implications from our sample is that the velocity centroid
pdf's in star forming regions are expected to drastically deviate from
the Gaussian shape predicted from hydrodynamic turbulence models. This
is indeed observed (\cite{lis98}, \cite{miesch98}). In our models this
is due to the presence of self-gravity inducing strong contracting
motions and clumpiness. However, other effects may lead to quite
similar distortions as well (see Klessen 2000 for more details).

\subsection{Evolution of clumps and cores}
\label{subsec:evolution}

\subsubsection{Clump mass spectrum}
\label{subsubsec:clump-mass-spectrum}

The structure of molecular clouds is extremely complex and
observations reveal a network of intersecting filaments and clumps on
all scales (e.g.~\cite{bal87,falg92,wise96}). Molecular clouds may be fractal and
various studies of the mass spectrum $N(m)$ of molecular clouds and of the
gas clumps inside of clouds indicate that their distribution may be
approximated by a power law of the form $dN/dm \propto m^{\alpha}$
with exponent $\alpha \approx -1.5$ (see e.g.~\cite{lor89,stu90,willi94,oni96,kram98,heithausen98}). This universal law is an important
constraint and test for models of molecular cloud
evolution. Numerical simulations {\em must} be able to reproduce this
structural feature of observed clouds. Indeed, our isothermal gas models
naturally lead to a power-law clump mass spectrum which is a
result of the interaction between gravity and gas pressure.

Representative for the entire set of isothermal larger-scale collapse
calculations, Fig.~\ref{fig:clump-spectrum-Z} shows the mass
distribution of clumps and condensed cores at four different stages of
the dynamical evolution of the high-resolution model $\cal I$, namely
initially and when 10\%, 30\% and 60\% of the available gas has been
accreted onto protostellar cores.  For each of these times the upper
panels compare the mass distribution of detected gas clumps (thin
line) with the observed clump mass distribution $dN/dm \propto
m^{-1.5}$, which translates into a slope of $-0.5$ when using
logarithmic mass bins (dashed line).  To identify
individual clumps, we have developed a clump-finding algorithm similar
to that of Williams et al.~(1994), but based on the framework of
SPH. For details see appendix \ref{app:clumps}. The thick line depicts the
mass distribution of condensed protostellar cores that have formed
within the more massive gas clumps (a detailed analysis of this
process is given in Sec.~\ref{subsubsec:growth-protostellar-cores}).
The lower panels show the position of each gas clump in a
mass--density diagram. Clumps without cores are shown as crosses.
Clumps containing one single core are denoted by open circles, those
with multiple cores by filled circles. Unresolved, very condensed
cores are plotted as stars with the density of cores being defined as
the core mass divided by the volume within the fixed accretion
radius. Therefore, they all fall along a straight line with slope
$1/3$.  The isothermal Jeans mass as function of density is indicated
by the diagonal line which separates the diagram into two regions.
Clumps that lie to the right exceed their Jeans mass and are due to
collapse, whereas clumps to the left are stabilized by gas pressure.
The vertical line indicates the SPH resolution limit
(Sec.~\ref{subsec:resolution-limit}). Clumps to the right of this line
are well resolved, objects to the left are ill-defined and may be
spurious results of the clump find algorithm. 

The clumping properties in all simulations, from model $\cal A$ with
$50\,000$ SPH particles to model $\cal I$ with ten times more
particles, are remarkably similar. This suggests that the dominant
dynamical processes are well treated and resolved. In all simulations,
the clump spectrum of the initial gas distribution {\em cannot} be
described by a simple power-law, it reflects the structural properties
of the Gaussian random field from which the initial conditions are
generated.  The initially linear growth of density perturbations is
not able to generate a hierarchical structure on all scales. This
requires considerable non-linear gravitational evolution to take
place. In the subsequent self-consistent dynamical evolution, the
clump distribution quickly achieves a universal mass distribution with
a power-law slope which is in excellent agreement with the observed
exponent $\alpha \approx -1.5$ (dashed line).  The core distribution,
on the other hand, deviates significantly from a power-law
distribution for smaller masses. Note, that the relative
underabundance of low-mass cores with respect to low-mass clumps is
not a resolution effect (see the vertical thin lines). All cores are
well resolved in our models.  Considerable deviations from a simple
power-law spectrum of clump masses occur again in the very late phases
of the evolution when most of the material is accreted onto dense
cores and the gas reservoir becomes significantly depleted. Then, huge
voids of very low density open up and the entire system more strongly
resembles a (proto)star cluster than a molecular cloud.  Within these
two extreme phases of the isothermal gas evolution, from the linear
initial gas distribution to the final star cluster, the complex
interplay between gravity and gas pressure naturally creates a
hierarchical structure with a distribution of clumps that is best
described by a simple power-law, in agreement with the
observations. As a reminder, when scaled to low density molecular
cloud regions with densities $n({\rm H_2})  \approx 10^2\,{\rm
cm}^{-3}$ (e.g.~Taurus) the total mass, $M=1$, of the system corresponds to
$6\;\!300\,$M$_{\odot}$. When considering high density regions with
$n({\rm H}_2) \approx 10^5\,{\rm cm}^{-3}$ where clusters of stars can
form (e.g.~the BNKL-region in Orion) then the physical mass of our
simulation is $200\,$M$_{\odot}$ (see Sec.~\ref{subsec:conversion}).

The lower panels in Fig.~\ref{fig:clump-spectrum-Z} indicate that
initially only very few high-mass clumps exist which exceed the local
Jeans mass.  These clumps collapse rapidly and form the first
condensed cores in their central region (see also
Sec.~\ref{subsubsec:ind-clumps}). During the subsequent evolution the
number of Jeans-unstable clumps and subsequently of protostellar cores
increases, populating a larger region to the right of the diagonal
line. If clumps merge which already contain protostellar cores, the
new more massive clump will contain multiple cores in its interior.
Competitive accretion (Price \& Podsiadlowski 1995; Bonnell et
al.~1997) and the gravitational interaction between the cores has
important consequences for the further evolution. It influences their
gas accretion and thus the final mass spectrum (more details are
discussed in Sec.~\ref{subsubsec:dyn-WW}). These often unpredictable,
probabilistic processes, resulting from chaotic multiple $N$-body
interactions of cores, are responsible for the difference in the
distribution of clump masses and of core masses. Whereas gas clumps
evolve according to the laws of hydrodynamics, cores behave like
accreting and gravitationally interacting $N$-body system.  As a
result, the mass spectrum of protostellar cores is well approximated
by a log-normal functional form whereas the clump spectrum shows a
power-law distribution.

\subsubsection{Formation and mass growth of cores}
\label{subsubsec:growth-protostellar-cores}
The location and the time at which protostellar cores form, is
determined by the dynamical evolution of their gas cloud. Besides
collapsing individually, clumps stream towards a common center of
attraction, where they may merge with each other or undergo further
fragmentation. As can be seen in Fig.'s \ref{fig:3D-cube-A} and
\ref{fig:3D-cube-I}, isothermal models evolve into a network of
intersecting sheets and filaments. The gas density is highest along
filaments and at their intersections. These are the locations where
dense cores form predominantly. The time of their formation in the
centers of unstable clumps depends strongly on the relation between
the time scales for individual collapse and fragmentation and
clump-clump merging.  When clumps merge that already contain cores
the dynamical interaction between the cores becomes important.

As illustration, we show in Fig.~\ref{fig:accretion-history} the
accretion history for a number of selected protostellar cores (a) from
model $\cal A$ and (b) from model $\cal I$.  The objects are numbered
chronologically according to their time of formation. Model $\cal A$
forms altogether 60 cores and model $\cal I$ 55. The figure shows the
following trends:

(a) The cores which form first tend to have the largest final
masses. They emerge from clumps that were present from the beginning
with the largest masses and highest densities and which are identified
with the most significant peaks in the fluctuation field. Jeans
unstable clumps with these qualities have very short collapse time
scales. Hence, these clumps quickly form a single cores which
are likely to accrete a large fraction of the parental gas before
the dynamical evolution of their environment changes the clump
properties too much, i.e.~before the clumps get dispersed or merge
with other clumps.  If a clump which already contains a core merges
with a non-collapsed gas clump, for example when streaming along a
filament towards a common center of attraction, then also the new
clump is likely to become completely accreted onto these cores.  Since
the first cores form in the highest-density regions more or less
independent of each other, their mass growth rate is dominated by the
matter originating from their vicinity.

(b) On the other hand, matter that contracts into dense cores at later
times ($t\gtrsim2$) has already undergone considerable dynamical
evolution. Clumps that were initially not massive enough to collapse
onto themselves merge until enough mass is accumulated for them to
collapse and build up a new protostellar core.  This needs time. For
those cores that form predominantly in the late stages of the
dynamical evolution the available gas reservoir is already
considerably depleted. Therefore, their average mass is rather
small. The final mass of the protostellar cores (i.e.~at the time when
90\% of the available gas mass has been accreted) for the standard
model $\cal A$ and the high-resolution model $\cal I$ in order of
their formation is plotted in Fig.~\ref{fig:sink-masses}.

Another aspect of the accretion process onto individual cores is
illustrated in Fig.~\ref{fig:mass-region}.  For the first nine cores
(Fig.~\ref{fig:mass-region}a) and for the last nine cores
(Fig.~\ref{fig:mass-region}b) that form in the high-resolution model
$\cal I$, it plots the average initial distances at $t=0.0$ between
the accreted SPH particles and the particle that gets converted into
the sink particle during the course of the simulation. This shows the
volume from which particles accrete onto protostellar cores. As
indicated above, the cores in Fig.~\ref{fig:mass-region}a accrete most
particles and the bulk of their final mass from their vicinity,
i.e.~from a distances less than $\sim 0.5$. With the total size of the
simulated cube being $2^3$, this corresponds to roughly $1/64$ of the
total volume. If the material would be randomly sampled from a
homogeneous cube of size $2^3$, the average distance would be $\sim
1.3$.  On the other hand, the smaller cores in
Fig.~\ref{fig:mass-region}b consist of matter that originates from a
larger volume, much closer to the value for random sampling. This
indicates that these clumps accrete from matter that already has
undergone considerable dynamical evolution and is well mixed.

Further information about the processes determining the formation of
and accretion onto dense cores is given in
Fig.~\ref{fig:clump-accretion}. It plots the contributions from
individual clumps to the final mass of selected cores.  As a general
trend, cores that form very early can accrete a large fraction of
their parental and neighboring clumps before clump interaction and
merging becomes important. On the other hand, matter that builds up
protostellar cores at the late stages of the dynamical evolution has
participated in large-scale motions and successive clump merging. It
is well mixed and the initial clump assignment is no longer
significant. Furthermore, the competitive accretion amongst groups of
cores in the interior of multiple merged clumps becomes more important
with the progression of time.  This phenomenon also contributes to the
distribution of available gas onto a larger number of
cores. Therefore, the fraction $\left \langle f_j \right \rangle$ of
material of individual clumps that gets accreted onto the cores
decreases with time.

\subsubsection{The Importance of dynamical interaction processes}
\label{subsubsec:dyn-WW} 
Besides the effects discussed in the previous section, the growth rate
of protostellar masses is strongly affected by the dynamical
interaction between the cores themselves. This phenomenon is closely
related to the competitive accretion of multiple cores within one
common gas envelope (\cite{price95}, \cite{bonnell97}) and becomes
important when clumps merge that already contain dense cores.  In the
center of the larger merged clump protostellar cores interact
gravitationally with each other.  Like in dense stellar clusters,
close encounters may lead to the formation of unstable triple or
higher-order systems and alter the orbital parameters of cluster
members. During the decay of unstable subsystems, protostellar cores
can get accelerated to very high velocities and may leave the
cluster. During this process, the less massive cores are more likely
to become ejected. Protostellar cores that get expelled from their
parental clump are suddenly bereft of the massive gas inflow from
their  collapsing surrounding. They effectively stop accreting
and their final mass is determined. The dynamical interaction between
cores is an important agent in shaping the mass distribution.

As an example, we discuss the history of protostellar cores in
the standard model. From Fig.~\ref{fig:accretion-history}a it is evident
that core \#5 stops accreting at $t\approx 0.8$, long before the
overall gas reservoir is depleted. The same applies to core \#7 at
$t\approx 0.9$. In both cases, the objects were involved in a 3-body
encounter that resulted in the expulsion from their gas rich parental
clumps. Fig.~\ref{fig:trajectory} depicts the trajectories of the cores
\#3, \#5, \#11, \#31, and \#46. At $t \approx 0.3$ core \#3 forms
within an overdense region and slightly later cores \#5 and \#11 form
in its vicinity from other Jeans-unstable density fluctuations. Their
parental gas clumps merge and the whole systems flows towards a local
minimum of the gravitational potential. The three cores soon build an
unstable triple system, continuing to accrete from the converging gas flow they
are embedded in (the detailed trajectories are shown in
Fig.~\ref{fig:trajectory}b, note the larger scale). At $t\approx 0.8$,
core \#5 is expelled and the remaining two cores form a wide binary,
which at $t \approx 1.6$ suddenly hardens due to the gravitational
encounter with core \#46 (see Fig.~\ref{fig:trajectory}c). Also this
hard binary is transient, the interaction with core \#31 pumps energy
into its orbit, and in the subsequent evolution, the orbital period
increases further during the encounter with a dense gas filament whose
tidal influence finally disrupts the binary at $t\approx 2.4$.  At
that time, accretion stops.

As another example from the high-resolution simulation,
Fig.~\ref{fig:accretion-history}b shows that at $t\approx 1.8$, core
\#19 stops accreting. It is thrown out of a dense clump at the
intersection of two massive filaments by a triple interaction with
cores \#1 and \#17.  However, it  remains bound to the gas clump
which grows in mass due to continuous infall and falls back into the
clump and resumes accreting at $t\approx2.0$. Cores \#9 and \#41 are
also expelled from their parental clumps but, unlike core \#17, their
accretion is terminated completely.  These dynamical interactions
between protostellar cores influence their mass distribution
significantly. In reality, ejected protostars may travel quite far and
could explain the extended distribution of weak-line T Tauri stars
found via X-ray observations in the vicinities of star-forming
molecular clouds (e.g.~\cite{neu95}, \cite{sterzik95}, \cite{kraut97},
\cite{wich97}).

As in our numerical scheme cores (i.e.~sink particles) only interact
gravitationally with each other, possible hydrodynamic processes are
suppressed. Dense cores that come close to each other exert mutual
tidal torques, they may merge or on the contrary strip off matter and
loose mass during this process. However, we estimate these effects to
be small. Matter accreted onto core particles would continue to
collapse inwards to form a young stellar object surrounded by a disk
on a very short time scale. The boundary introduced by the sink
particle is determined by the resolution limit of the code. Scaled to
physical units the diameter of the core particle is roughly 600 AU in
case of high densities of $n({\rm H}_2) \approx 10^5\,{\rm cm}^{-3}$
(as in the Orion star forming region) and about $10\,000$ AU in low
density regions like Taurus (using $n({\rm H}_2) \approx 10^2\,{\rm
cm}^{-3}$ and $T\approx10\,$K). Matter inside the core is always
expected to be strongly concentrated in the protostar in the
center. The necessary cross section for merging or significant tidal
perturbation (e.g.~Hall, Clarke \& Pringle 1996) is therefore
typically much smaller than the size of the sink particles. To
estimate how possible core mergers could influence the conclusions
derived in this paper we took the initial conditions of simulation
$\cal A$ and repeated the calculation allowing for core particles to
join together. We studied three different cases. The most stringent
one was that we merged sink particles if they overlap by at least
90\%, then if they overlap by 10\% or more, and finally if two sink
particles just touch each other.  The resulting shape and width of the
final core mass distribution remained essentially unchanged in all
three cases compared to simulation $\cal A$, only the number of cores
decreased.

\subsection{Properties of (individual) gravitationally unstable clumps}
\subsubsection{Properties of individual clumps}
\label{subsubsec:ind-clumps}
Whereas the previous section discussed global features of the clump
and core distribution, this section investigates in detail the
properties of individual clumps. This is important for studies of
the collapse of single clumps and their fragmentation into binaries
and multiple stellar systems
(\cite{burkert93}, \cite{burkert96}, \cite{boss97}, \cite{burkert98},
\cite{burkert98b}).  The
simulations presented here are no longer able to resolve the collapse
and sub-fragmentation of individual protostellar cores once they have
been substituted by sink particles.  To study this subsequent
evolution, a new simulation just concentrating on the collapse of one
single core would be required. To connect the final stages of the
larger-scale simulations discussed here and the initial conditions for
detailed collapse calculations, knowledge about the properties of
individual cores is necessary. Important parameters are their density
structure and their rotational properties.  We note, however, that gas
clumps are never isolated.  Interactions with the environment are very
important for the mass spectrum, the spatial distribution and the
geometrical shape of clumps and cores.

The clumps generally are highly distorted and triaxial.  Depending on
the projection angle, they often appear extremely elongated, being
part of a filamentary structure which may appear as a chain of
connecting, elongated individual clumps. More complicated irregular
shapes are also common which result from recent clump mergers at the
intersections of filaments. A selection of clump shapes is presented
in Fig.~\ref{fig:ind-clumps} which plots the contour lines for eight
high-density clumps in simulation $\cal I$ at time $t=1.2$ projected into the
$xy$-, $xz$- and $yz$-plane.  At this state of the dynamical
evolution, four collapsed cores have formed which accreted 1\% of the
total gas mass. This time is most appropriate to determine clump
properties, because the system has already undergone substantial
evolution (the power-law clump spectrum is well established) but most
of the gas has not yet been accreted onto protostellar cores.  The
clumps that already formed dense cores in their interior are plotted
in the first part of the figure, the second part shows clumps
whose central densities are not yet high enough to form a
condensed core. The clumps are numbered according to their peak
density, i.e.~clump \#1 has the highest central density. The (surface)
density contours are spaced logarithmically with two contour levels
spanning one decade, $\log_{10}\Delta \rho = 1/2$. The lowest contour
is a factor of $10^{1/2}$ above the mean density $\left \langle \rho
\right \rangle = 1/8$. The black dots indicate the positions of the
dense protostellar cores. Note the similarity to the appearance of
observed dense (pre-stellar) clumps (for instance Fig.~1 in
\cite{myers91}). It is clearly visible that the clumps are very
elongated. The ratios between the semi-major and the semi-minor axis
measured at the second contour level are typically between 2:1 and
4:1. However, there may be significant deviations from simple triaxial
shapes, see e.g.~clump \#4 which is located at the intersection of two
filaments. This clump is distorted by infalling material along the
filaments and appears `y'-shaped when projected into the
$xz$-plane. As a general trend, high density contour levels typically
are regular and smooth, because there the gas is mostly influenced by
pressure and gravitational forces. 
On the other hand, the lowest level samples gas that is
strongly influenced by environmental effects. Hence, it appears patchy
and irregular. The location of the condensed core within a clump is
not necessarily identical with the center-of-mass of the clump,
especially in irregularly shaped clumps.

Typically, the overall density distribution of identified clumps in
our simulations follows a power law and the density increases from the
outer regions inwards to the central part as $\rho(r) \propto
1/r^2$. For clumps that contain collapsed cores, this distribution
continues all the way to the central protostellar object. However, for
clumps that have not yet formed a collapsed core in their center, the
central density distribution flattens out. As an example,
Fig.~\ref{fig:radial-density-profile} plots the averaged radial
(surface) density profile of the $xz$-projection of clump \#4 and of
clump \#12.  This is in accordance with analytical models of
isothermal collapse (see \cite{lars69}, \cite{penston69}; and also
\cite{whit85}) and with the observational data for dense cores in
dark molecular clouds (see e.g.~\cite{myers91}, \cite{ward94},
\cite{chini97}, \cite{motte98}).

\subsection{Properties of the evolving protostellar cluster}
\subsubsection{Binarity and clustering properties of protostellar cores}
\label{subsubsec:clustering}
An important statistical quantity derived from observational data is
the 2-point correlation function or equivalently the mean surface
density of companion stars $\xi$ as function of separation $r$.
Larson (1995) found that in the Taurus star-forming region the mean
surface density of companions follows a power law as function of
separation with a break in the slope at the transition from the binary
to the large-scale clustering regime. It occurs at separations for
which the binary system blend into the background distribution of
cluster stars.  This analysis has also been applied to other
star-forming regions (\cite{simon97}, \cite{bate98},
\cite{gladwin98}) . Larson identified the transition separation with
the typical Jeans length in the molecular cloud. For Taurus this may
be true.  In Orion this conclusion fails. In general, it can be shown
that the transition separation depends on the volume density of stars,
the extent of the star-forming regions along the line-of-sight, the
volume-filling nature of the stellar distribution and on the details
of the binary distribution (\cite{bate98}).  Additionally, the
transition separation evolves in time due to dynamical interactions
amongst the cluster members.  Note that the 2-point correlation
function is never unique. It cannot differentiate between hierarchical
(fractal) and non-hierarchical structure, and different stellar
distributions may lead to the same 2-point correlation
function. Therefore, the additional analysis of the spatial
distribution by eye is essential for a meaningful interpretation of
$\xi$. Alternatively, to obtain quantitative results, the calculation
of higher-order correlation functions may be useful, which is common
in studying the large-scale structure of the universe
(\cite{peebles93}).  However, for star-forming regions this has never
gone beyond the second order terms. Hence, we restrict our current
analysis to a discussion of the 2-point correlation function $\xi$ of
the spatial distribution of protostellar cores (sink particles).

Figure~\ref{fig:larson-X} plots the mean surface density of companions
$\xi$ as function of separation $r$ for the protostellar cluster that
forms in the standard model $\cal A$ at four different times of its
evolution.  At each point in time, the left panel gives the spatial
distribution of the protostellar cluster projected in the $xy$-, $xz$-
and $yz$-plane. The resulting functions $\xi$ for each projection are
shown in the right panel.  We use data from model $\cal A$ because it is the
model which we advanced furthest in time. However, the conclusions
apply to all models discussed in this paper: The initial binary
fraction is very large. Roughly 75\% of the cores are part of a
(bound) binary or higher-order system which dominate the function
$\xi$ at small separations $r$. At larger $r$ the function $\xi$
flattens out which reflects the fact that the binary systems are
relatively homogeneously distributed throughout the entire volume
(Fig.~\ref{fig:larson-X}a).  The value of
the plateau of $\xi$ indicates the projected density of this
homogeneous distribution.

At the intersections of filaments clumps which already contain one or
more cores merge, forming larger clumps which subsequently contain a
cluster of cores. As a result of this process, in
Fig.~\ref{fig:larson-X}b one sees several small aggregates distributed
throughout the simulated volume. Within these small dense clusters
complex dynamical interaction between cores takes place (cf.~with
Sec.~\ref{subsubsec:dyn-WW}) leading to the destruction of soft
(i.e.~weakly bound) binaries. At the same time hard binaries become
even harder (i.e.~more strongly bound) and the function $\xi$ extends
to smaller spatial separations. However, due to this secular evolution
the overall binary fraction decreases.  During the progression of the
evolution, the small protostellar aggregates follow the streaming
motion of their surrounding gas envelopes and merge, thereby forming a
single bound cluster (see Fig.~\ref{fig:larson-X}c). The protostellar
system has reached a state of minimum spatial extent. The overall size
of the cluster is reflected by the steep decline of the function $\xi$
at large separations.

At that stage about 90\% of the available gas has been
accreted and the subsequent evolution of the cluster is almost
entirely determined by gravitational $N$-body interactions. The cluster
re-expands and rapidly develops a core/halo structure which is typical
for collision-dominated self-gravitating $N$-body systems (see
e.g.~\cite{BT87}). This process is expedited by the fact that cores
which have been accelerated to high velocities by close encounters
with hard binaries cannot leave the simulation box due to the periodic
boundary conditions.  Once those `bullets' have trespassed the
boundaries of the cube, they are reinserted at the opposite side with
the same velocity and may again interact with other cluster members to
transfer energy to more slowly moving cores. In reality they would be
ejected from the cluster and leave the star-forming cloud.  At these
late stages of the evolution, a clear change in the slope of the
function $\xi$ can be seen at the break between the binary regime and
the large-scale clustering (see Fig.~\ref{fig:larson-X}d). The
fraction of fast moving ejected cores has grown to roughly 50\%. At
the same time, the binary fraction has decreased to 15\%.

Duquennoy \& Mayor (1991) analyzed a sample of nearby G-dwarfs and
found that the distribution of orbital periods of binary stars in
their sample follows approximately a broad Gaussian with a peak at
$\sim 10^{4.8}$ days which corresponds to a separation of $\sim
32\;$AU (using Kepler's third law and assuming the typical mass of a
G-dwarf is $\sim 1\,$M$_{\odot}$). From the broad distribution of
orbital periods it follows immediately that the function $\xi$ of mean
surface density of binary companions decreases with separation as
$1/r^2.$\footnote{A broad distribution of orbital periods (in
$\log_{10} P$) is equivalent to a broad distribution of binary
separations $\log_{10} r$. To the lowest order it can be approximated
as being flat, i.e.~$dN/d\log_{10} r = K$. Then, $dN = K/r\;dr$ is the
number of binary companions with separations in the range $r$ to $r +
dr$ and the mean surface density of companions in the 2-dimensional
projection follows as $\xi(r) = N^{-1}dN/(2\pi r dr) = N^{-1}K/ (2\pi
r^2) \propto 1/r^2$.} The same behavior is found in our
systems. Figure \ref{fig:binary-separations} illustrates the
distribution of semi-major axes for the identified {\em bound} pairs
of cores in simulation $\cal A$ at selected times. For each core its
closest neighbor is found and if the pair is bound (i.e.~if the sum of
relative kinetic and potential energy is negative) the kinematic
data at the present time are used to calculate the orbital
parameters. In case of isolated pairs this works well, however, most
cores are part of higher-order systems. In this case the given value
is only a rough estimate of the true orbital characteristics as in
unstable and highly chaotic few-body system well defined orbital
parameters do not exist. We find that the logarithmic distribution of
binary separations is broad and relatively flat and correspondingly
that the function $\xi$ falls off as $1/r^2$ in the binary regime
(Fig.~\ref{fig:larson-X} at small $r$).  When compared with the
observations it should be noted furthermore that the systems of
protostellar cores in our simulations correspond to wide
binaries. Close binary system would form from {\em sub}-fragmentation
of individual protostellar cores (e.g.~Burkert \& Bodenheimer 1996).
This is a process which we cannot treat in the present numerical
scheme. However, as indicated in Fig.~\ref{fig:ang-momentum-2}, also
the angular momenta of individual protostellar cores in our models
follow a relatively flat and broad logarithmic distribution spanning
at least three orders of magnitude.  If we assume that a rotating
protostellar core typically breaks up into a binary system with
similar angular momentum the period distribution of close binaries
should also be broad, as observed.  This result follows from the
stochastic processes which determine the formation and evolution of
protostellar cores. They influence the properties of binary systems
and at the same time also affect the angular momentum distribution of
individual cores.

\subsubsection{Rotational properties of protostellar cores}
\label{subsubsec:rotational-properties} 
A very important parameter for core collapse and possible
sub-fragmentation is their specific angular momentum. 
Conservation of angular momentum prevents material from the envelope
of a star forming core
to directly accrete onto the central protostar. Only the inner part 
is able to immediately fall onto the star, the bulk of the
infalling envelope accumulates in a rotationally supported disk around
the central object.  Angular momentum is transferred outward on 
viscous time scales $\tau_{\rm vis}$, by this moving matter inwards
towards the central star. Typically, $\tau_{\rm vis}$ is by a factor of
$10\;$--$\;100$ larger than the rotational time scale which is
comparable to the free-fall time scale $\tau_{\rm ff}$ (see
e.g.~\cite{pringle81}).

These processes take place deep inside the protostellar cores, hence they
cannot be resolved by our numerical scheme. However, we can keep track of the
the total angular momentum accreted onto each condensed core in our
simulation. Figure~\ref{fig:ang-momentum-1} plots the time evolution of the
angular momentum vector $\vec{L}$ of all protostellar cores in the
high-resolution model $\cal I$. The left panel describes the evolution of each
individual component of the vector. The right panel plots the distribution of
$L_x$, $L_y$ and $L_z$ at the end of the simulation at $t=3.9$. For
comparison, we overlay the final distribution (at $t=5.6$) of the angular
momenta of the cores in the standard model $\cal A$ with dashed lines. Both
distributions are statistically indistinguishable. From the plots, we see that
the evolution of the angular momentum of individual cores can be very complex
and intimately reflects the rotational properties of the environment they are
embedded in: the angular momentum of the protostellar core is determined by
the angular momentum of the clump it forms in. Clumps that merge at the
intersection of two filaments can accumulate considerable angular momentum
which is transfered onto the core by accretion. The angular
momentum vector of cores may even change its sign if material is accreted
which rotates counterclockwise with respect to the core.

The distribution of the absolute values of angular momenta
$|\vec{L}\;\!|$ of the protostellar cores in simulation $\cal I$ is
plotted in Fig.~\ref{fig:ang-momentum-2}a.  Values are shown for
$t=2.8$ (solid line), i.e.~when 60\% of the gas is converted into
dense cores, and at the final stage, $t=3.9$ (dashed line). As already
indicated in Fig.~\ref{fig:ang-momentum-1}, the angular momenta are
very broadly distributed and peak at $\sim 10^{-5}$. Note again that
all values are given in dimensionless units with length and mass
scales set to one. Figure~\ref{fig:ang-momentum-2}b plots the
distribution of the specific angular momenta $|\vec{L}\;\!|/m$ of the
cores, again at $t=2.8$ (solid line) and at $t=3.9$ (dashed line).
The mass dependence of $|\vec{L}\;\!|/m$ is shown in
Fig.~\ref{fig:ang-momentum-2}c. There are no massive cores with low
specific angular momentum. However, there is no clear correlation
between both quantities.  

Using the scaling properties discussed in
Sec.~\ref{subsec:conversion}, the dimensionless $|\vec{L}\;\!|/m$
values can be converted into physical units. When we apply our model
to the high-density regime (e.g.~to the BN region in the Orion
molecular cloud with $n({\rm H}_2) \approx 10^5\,{\rm cm}^{-3}$) then
$|\vec{L}\;\!|/m=1$ corresponds to $10^{23}\,{\rm cm}^2{\rm
s}^{-1}$. In the case of low densities of $n({\rm H}_2) \approx
10^2\,{\rm cm}^{-3}$ (the average value in Taurus) this converts into
$3.7\times 10^{24}\,{\rm cm}^2{\rm s}^{-1}$. The specific angular
momenta of protostellar cores in our simulations are therefore of the
order of $10^{20}$ to $10^{21}\,{\rm cm}^2{\rm s}^{-1}$ and agree
remarkably well with the observed values for wide binaries and
protostellar objects (e.g.~Bodenheimer 1995).

When looking at the spatial distribution of the angular momentum
vectors of protostellar cores, there is a correlation between the
location and the orientation.  As is visible in
Fig.~\ref{fig:ang-momentum-3}, the angular momentum vectors of cores
that form in the same region tend to be aligned.  These cores form
from gas that has similar global flow patterns. In clumps which
contain multiple cores as the result of merging, all cores accrete
from a common environment with a certain well defined angular momentum
vector. As a result, the rotational properties of these cores tend to
be comparable. In some star-forming regions, molecular outflows from
young stellar objects indeed appear to be correlated and aligned
(e.g.~in the northern part of the L1641 cloud, see Fig.~14 in the
review article by \cite{reip89}). Assuming that protostellar outflows
are associated with the rotational properties of the protostellar
object they emerge from, this implies a correlation between the
angular momenta of the protostellar cores similar to the one found in
our numerical models.  On the other hand, in other regions no
correlation is found at all (see Eisl{\"o}ffel \& Mundt 1997, Stanke,
McCaughrean \& Zinnecker 1998, Reipurth, Devine \& Bally 1998 for
recent surveys). Taking all together, the observational data are not
conclusive.

\subsubsection{Boundedness of protostellar clusters}
\label{subsubsec:bound}
In the simulations described in this paper, the dynamical evolution of
the gas results in the formation of dense clusters of collapsing cores
that form protostars (see Fig.'s~\ref{fig:3D-cube-A} and
\ref{fig:3D-cube-I}, and the left panels of
Fig.~\ref{fig:larson-X}). These clusters are bound throughout their
entire evolution. To specify this in more detail,
Fig.~\ref{fig:boundedness}a plots for the protostellar clusters
forming in simulation $\cal A$ (dashed lines) and $\cal I$ (solid
lines) the time evolution of the total kinetic energy subdivided into
the contribution from the internal velocity dispersion (thick line)
and from the center-of-mass motion (thin line). The kinetic energy is
almost entirely dominated by the internal random motions of the
cluster members. In Fig.~\ref{fig:boundedness}b, the evolution of the
potential energy is given. Only the gravitational interaction between
cores themselves is taken into account. The potential of the gas in
which the cluster is embedded is excluded in order to estimate whether
the cluster would dissolve in case of a sudden removal of the gas.
The cluster in the high-resolution model $\cal I$ (solid lines) is
still contracting when the simulation stops at $t=3.9$.  The cluster
in model $\cal A$ is allowed to evolve further and develops the
typical core/halo structure of collisional $N$-body systems. The
global virial coefficient $\eta_{\rm vir} \equiv 2E_{\rm int}/|E_{\rm
pot}|$ is plotted in Fig.~\ref{fig:boundedness}c. Once the clusters
have formed they remain marginally bound, i.e.~$\eta_{\rm vir}
\lesssim 1$, even if all gas is removed. These conclusions do not
change when taking into account only the cores that lie within the
half-mass radius of the clusters.  The cumulative mass of the clusters
is given in Fig.~\ref{fig:boundedness}d.  For $t \geq 2.0$ more than
50\% of the total mass is condensed into protostellar cores (sink
particles). It is, however, interesting that already much earlier,
during the formation of the cluster $\eta_{\rm vir}$ becomes constant
with $\eta_{\rm vir} \approx 0.7$.  Dense protostellar cores tend to
form such that the increase in the absolute values of the potential
energy of the stellar cluster neglecting the potential of the gas
component is always more or less balanced by an increase of its
kinetic energy, even at very early times when the gravitational
potential is strongly dominated by the gas.  The 3-dimensional
velocity dispersion and the line-of-sight velocity dispersion of the
two clusters as function of time is described in
Fig.~\ref{fig:velocity-dispersion}. Again the dashed line denotes
model $\cal A$ and the solid line model $\cal I$.  The fact that in
our models the line-of-sight measurements along the different axes are
almost identical to each other demonstrates that the cluster is
kinematically well mixed and isotropic.  When scaled to low-density
star-forming regions like Taurus, a dimensionless value of $\sigma=1$
corresponds to $2.2\,$km$\,$s$^{-1}$.  In the case of a high-density
region, this corresponds to $3.0\,$km$\,$s$^{-1}$ (see
Sec.~\ref{sec:scaling}). The values calculated from our simulations
when roughly 50\% to 60\% of the gas has been accumulated in cores ($t
\approx 2$) are in agreement with the measurements in Taurus and in
the Trapezium cluster in Orion which both have comparable velocity
dispersions of $\sigma \approx 2.5\,$km$\,$s$^{-1}$ (for Taurus see
\cite{frink97}, and for Orion see \cite{jones88}).  However, for very
high core-formation efficiencies, when more than 60\% of the gas has
been converted into condensed cores, the derived velocity dispersions
appear to be too high. In order to reproduce the observed velocity
dispersions we predict core formation efficiencies of order 50\% to
60\%. This is reasonable as one would expect that energetic feedback
processes between the newly formed stars and their gaseous environment
disperse gas clouds before they have completely condensed into stars.
Note that the total star formation efficiency in the considered
molecular cloud region is the product of the core formation efficiency
and the efficiency with which the material accumulated in dense cores
accretes onto their central stellar objects. The latter one is
expected to be close to unity (for detailed simulations see Wuchterl
\& Tscharnuter 2000; for observational evidence see Motte, Andr\'{e}
\& Neri 1998)

\subsection{Implication for the IMF}
\label{subsec:implications-IMF}
In this section, the time evolution and the properties of the mass
distribution of protostellar cores in our simulations and their
relation to the initial stellar mass function are discussed. As
analyzed in the previous sections, protostellar cores form in the
centers of Jeans-unstable massive gas clumps and grow in mass via
competitive accretion. This process is strongly influenced by the
presence of unpredictable dynamical events which determine the shape
of the mass spectrum. In Fig.~\ref{fig:clump-spectrum-Z} the mass
distribution of identified gas clumps and protostellar cores in the
high-resolution model $\cal I$ has been introduced at four different
stages of the dynamical evolution of the system, i.e. initially and
when 10\%, 30\% and 60\% of the available gas has been converted into
condensed cores. Whereas the mass spectrum of gas clumps is best
described by a power-law function, the distribution of core masses
follows a Gaussian.  Again for the high-resolution model,
Fig.~\ref{fig:mass-spectrum-1} plots the mass distribution of
protostellar cores at times when the cluster of cores has accreted a
total mass fraction of (a) $M_*=5$\%, (b) 15\%, (c) 30\%, (d) 45\%,
(e) 60\%, and (f) 85\%. Spanning two orders of magnitude, the mass
distribution of protostellar cores is very broad and peaks
approximately at the overall Jeans mass of the system, $\left \langle
M_{\rm J} \right \rangle = 1/222 = 10^{-2.3}$ (see
Sec.~\ref{sec:scaling}).  This is somewhat surprising, given the
complexity of the overall dynamical evolution.  The Jeans mass is a
function of density and may vary strongly for different clumps. In a
statistical sense, the system retains `knowledge' of its (initial)
average properties and the `typical' core mass is closely related to
the `typical' Jeans mass.  In the initial conditions the density
contrast is limited. Hence, the gravitationally unstable clumps which
in Fig.~\ref{fig:clump-spectrum-Z} are located to the right of the
tilted line indicating the Jeans limit as function of density, all
have densities comparable to the mean density of the system,
i.e.~their local Jeans mass is $\sim \left \langle M_{\rm J} \right
\rangle$. These clumps form the first generation of cores which will
become very massive and populate the upper end of the mass
spectrum. At later stages, initially smaller clumps have merged and
grown dense  enough to become Jeans unstable too.  They begin to
collapse and form cores preferably at the low-mass end of the
spectrum. As long as there is enough gaseous material, the growth rate
of already existing cores to larger masses and the formation rate of
new low-mass cores are comparable and the mass distribution evolves
symmetrically (Fig.~\ref{fig:mass-spectrum-1}a -- e). However, at very
late stages of the evolution, the distribution gets skewed towards
higher masses which is an effect of the depletion of the gas
reservoir. There is still mass available for accretion, but it is not
sufficient to form new cores (see Fig.~\ref{fig:mass-spectrum-1}a --
f). The fact that the total gas reservoir is limited not only modifies
the shape of the distribution, it also influences its peak value.  At
early stages, the median core mass is slightly below the average Jeans
mass, at late stages it lies above (see also
Tab.~\ref{tab:fitting}). Competitive accretion and the dynamical
interaction of protostellar cores as members of dense clusters (see
Sec.~\ref{subsubsec:dyn-WW}) do not alter the shape of the mass
distribution, but may widen it further.

Figure~\ref{fig:mass-spectrum-2} presents the mass spectra of
protostellar cores for {\em all} simulations discussed in this paper
(see Tab.~\ref{tab:models-N=2}) at the time when we would expect gas
to he heated and dispersed by newly formed stars (see section 6.4.3),
that is when roughly 60\% of the gas is accumulated in protostellar
cores. The distributions are very similar and the variations in widths
and centroid are small.  Since the data sets of observed protostellar
cores are not yet large and accurate enough to derive well established
mass distributions\footnote{For $\sim$ 60 pre-stellar cores (i.e.\ for
self-gravitating gas clumps without central collapsed object) in the
$\rho$-Ophiuchus cloud, a mass spectrum is presented by Motte et al.\
(1998). See their Fig.~5.}, we have to go one step further and compare
the core mass spectrum with the initial mass function of stars (IMF).
The present simulations cannot resolve the conversion of individual
protostellar cores into stars. Since detailed collapse simulations
show that perturbed cores tend to break up into multiple systems we
can only make predictions about the system mass function. We adopt the
IMF for multiple stellar systems (i.e.~without corrections for binary
stars and higher-order systems) from Kroupa et al.~(1990) which we
compare with our numerically obtained core mass distribution.  For
this comparison we need to know the efficiency of individual dense
cores to form stars inside. As their mass loss due to radiation, winds
and outflows is expected to be small, most of the material accumulated
in dense cores will be accreted onto the central protostars (Wuchterl
\& Tscharnuter 2000). Hence, we take the efficiency to convert dense
cores into stars to be of order unity. The overall star-formation
efficiency then depends only on the rate at which molecular gas forms
dense cores. It general, this rate evolves with time and is determined
by the properties of the underlying turbulent velocity field. This is
particularly important in the case of `isolated' star formation
(Klessen, Heitsch \& Mac\ Low 2000). Here, we concentrate on the
`clustered' mode of star formation and investigate molecular cloud
regions where turbulence has decayed completely. The overall
efficiency may also be influenced by the presence of massive O and B
stars in the vicinity or within of the considered molecular cloud
region. Once formed, their radiation would ionize the molecular gas
and could prevent subsequent core formation and growth.  As this effect is
not included in our present investigation, the overall core formation
efficiency therefore in principle remains a free parameter. 

To estimate the appropriate density scaling in our models
necessary for the computed and the observed mass distributions to
agree, we require that both distributions peak at the same mass.
Recall that the simulated core mass distribution reaches its maximum
roughly at the overall Jeans mass of the system.  To allow for a
statistically significant comparison with observational data, we merge
together the mass distribution of all nine models with $\alpha=0.01$
and $P(k) \propto 1/k^2$.  The resulting combined mass spectrum is
shown in Fig.~\ref{fig:mass-spectrum-3} at three stages of the
evolution, when the fraction of mass in protostellar cores is $M_*
=30$\%, $M_* = 60$\%, and $M_* = 90$\%, respectively.  The
best-fitting Gaussian representation of each distribution is plotted
using open circles. The location of the centers and the widths are
specified in Tab.~\ref{tab:fitting}. The log-normal description of the
observed IMF for multiple stellar systems (the MS model in Kroupa et
al.\ 1990) is indicated by the dashed lines.  We use the following
functional form,
\begin{equation}
\label{eqn:log-normal-function}
\xi(\log_{10}m) = \xi_0
\exp\left[-\frac{(\log_{10}m-\log_{10}\mu)^2}{2(\log_{10} \sigma)^2}\right]\,,
\end{equation}
where $N=\xi(\log_{10}m) d\log_{10}m$ is the number of objects per
logarithmic mass interval. The quantities $m$, $\mu$ and $\sigma$ are
given in units of the solar mass, $\log_{10}\mu$ and $\log_{10}\sigma$
determine peak and width of the distribution, and $\xi_0$ is the
normalization. With this functional form the observed IMF peaks at
$\log_{10} \mu = -0.31$ and has a width of $\log_{10} \sigma =
-0.38$. For the efficiency $M_* = 30$\%, the core mass distribution
has its maximum slightly below the average Jeans mass, $m_{\rm peak} =
0.86 \:\!\langle M_{\rm J}\rangle$. For agreement with the IMF the
system needs to be scaled such that the average Jeans mass is $\langle
M_{\rm J} \rangle = 0.58\,{\rm M}_{\odot}$. For a gas temperature
$T=10\,$K, this corresponds to a mean density $n({\rm H}_2) = 2.3
\times 10^5\,$cm$^{-3}$. For $M_* = 60$\% and $M_* = 90$\%, where
$m_{\rm peak} = 1.4\:\!\langle M_{\rm J}\rangle$ and $m_{\rm peak} =
2.0 \:\!\langle M_{\rm J}\rangle$, these values are $n({\rm H}_2) =
6.2 \times 10^5\,$cm$^{-3}$ and $n({\rm H}_2) = 1.3 \times
10^6\,$cm$^{-3}$, respectively (see Tab.~\ref{tab:fitting}). These
densities agree well with those observed in cluster forming regions
(see e.g.\ Motte et al.\ 1998 for the $\rho$-Ophiuchus cloud). Note,
that the width of the core mass distribution is slightly larger than
the one of the IMF of multiple stellar system by Kroupa et al.\
(1990), however, it is somewhat smaller compared to the original
Miller \& Scalo (1979) mass function (their model of constant
birthrate over $12\times 10^9$ years has $\log_{10} \sigma = -0.68$;
see the dotted lines in Fig.\ \ref{fig:mass-spectrum-3}).  Given the
uncertainties involved in the observational determination of the IMF
(e.g.~Scalo 1986, 1998a) the agreement is remarkable and the
significance of these deviations is low.

Our dynamical model of clustered star formation predicts a universal
initial mass function with a log-normal functional form similar to the
observational data for multiple stellar systems.  The overall nature
of the star formation process can only be understood in the frame work
of a statistical theory, where a sequence of probabilistic events may
naturally lead to a log-normal IMF (e.g.~Zinnecker 1990, Larson 1992,
1995, Richtler 1994, Price \& Podsiadlowski 1995, Murray \& Lin 1996,
Elmegreen 1997; also Adams \& Fatuzzo 1996). Since the final mass
distribution of protostellar cores in our self-gravitating, isothermal
models is a consequence of the chaotic dynamical evolution during
the accretion phase, our simulations support this hypothesis.

\section{Summary}
\label{sec:summary}

In this paper we numerically investigated the dynamical evolution and
fragmentation of molecular clouds and discussed the interplay between
gravity and gas pressure. We identified the processes that dominate
the formation and evolution of (proto)stellar clusters and determine
their properties. From the results we can conclude that even simple,
isothermal models of self-gravitating clumpy clouds are able to
explain many of the observed features of star-forming regions. This is
rather surprising, given the fact that magnetic fields and energetic
heating processes of newly formed stars could in principle be
important, both of which have been neglected.  At this point the
reader should be cautioned again that our isothermal models do not
include any feedback effects from the star formation process
itself. At late stages of the evolution energy and momentum input from
young stars is likely to modify the dynamical properties of the
accreting gas and the simple isothermal approach which we follow here
needs to be extended by more elaborate schemes.

Our simulations show that, in general, the formation of a cluster of
condensed cores and protostars through gravitational collapse and
fragmentation of a molecular cloud region is extremely complex. The
dynamical evolution of molecular gas is determined by the interplay
between self-gravity and gas pressure. This creates an intricate
network of filaments, sheets and dense clumps. Some clumps will become
gravitationally unstable and undergo rapid collapse. While contracting
individually to form protostellar cores in their interior, gas clumps
stream towards a common center of attraction: the dynamical evolution
of molecular clouds involves processes acting simultaneously 
on different length scales and time scales.
While following the large-scale flow pattern, gas clumps
can undergo further fragmentation or merge at the intersections of
filaments.  At that stage, the central regions of some clumps will
have already collapsed to sufficiently high densities to be identified
as protostellar cores. These cores rapidly grow in mass via accretion
from their parental gas envelope. When clumps merge, the
newly formed clump may contain a multiple system of
protostellar cores which subsequently compete with each other for
accretion from the same limited and rapidly changing reservoir of
contracting gas in which they are embedded. Since the cores are
dragged along with the global gas flow, a dense cluster of
accreting protostellar cores builds up quickly. Analog to dense stellar
clusters, its dynamical evolution is subject to the complex
gravitational interaction between the cluster members: close
encounters occur frequently and will drastically alter the orbital
parameters of cores. This leads to the formation of unstable triple or
higher-order systems, and consequently a considerable fraction of
protostellar cores becomes expelled from the cloud. These cores
effectively stop accretion and their final mass is determined.

The presence of unpredictable dynamical events in the overall gas flow
and the evolution of the nascending protostellar cluster very
efficiently erases the memory of the initial configuration.  For this
reason, we cannot predict the detailed evolution of individual objects
just from the initial state of the system. Only the properties of an
{\em ensemble} of protostellar cores, for example their kinematics and
mass distribution, can be determined in a probabilistic sense. A
comprehensive theory of star formation needs to be a {\em statistical
theory}. Some first attempts to formulate a statistical model of the
star-formation process appear very promising (e.g.~Zinnecker 1990,
Larson 1992, 1995, Richtler 1994, Price \& Podsiadlowski 1995, Murray
\& Lin 1996, Adams \& Fatuzzo 1996, Elmegreen 1997; for an overview
see Scalo 1998b) and are supported by the results of our numerical
study. Taken together, our simulations strongly suggest that {\em
gravitational} processes and accretion dominate the early phases of
star formation.

We extend the above overview by giving a detailed list of the features
and results of our calculations derived from the comparison of our
numerical models with specific observational properties of molecular
clouds and young stellar clusters:

\begin{itemize}

\item {\bf One-point probability distribution function of density and
line-of-sight velocity centroids:} During the dynamical evolution of
the system gravitational contraction on large scales and the collapse
of individual gas clumps result in considerable distortions of the
pdf's away from the initial Gaussian behavior. The pdf of the
logarithm of the density is best described by a (multiple) power-law
distribution and the  pdf of the line-of-sight velocity centroids
develop  extended high-velocity wings. This also holds for the
line profiles themselves (Sec.~\ref{subsec:pdfs}).

\item {\bf Clump mass spectrum:} During the early phases of
star formation our simulations of self-gravitating isothermal
gas are able to reproduce the observed power-law clump-mass spectrum
of molecular clouds.  This is due to the progression of non-linear
gravitational attraction and the disintegration of small clumps by gas
pressure. The observed spectrum is best fit at times between the
formation of the first condensed objects up to the time when depletion of
the gas reservoir becomes considerable. Neither the initial Gaussian
fluctuation spectrum, nor the final stages of the evolution when most
of the gas is condensed into protostellar cores, give a clump-mass
distribution which is in agreement with the observations.  This is discussed in
Sec.~\ref{subsubsec:clump-mass-spectrum}.  

\item {\bf The Shapes of Individual Clumps:} Dense Jeans-unstable gas
clumps are the precursors of protostellar cores. As being part of a
complex network of filaments, individual clumps are typically very
elongated objects with ratios between semi-major and semi-minor axis
of 2:1 to 4:1. However, in many cases (especially at the intersection
of two filaments) they can be quite irregularly shaped. These features
are in agreement with observed dense pre-stellar cores in dark clouds
(Sec.~\ref{subsubsec:ind-clumps}).

\item {\bf Formation and Growth of Protostellar Cores:} The formation
and growth of protostellar cores is subject to a progression of
statistical events. However, we can identify the following trends in
our models (Sec.~\ref{subsubsec:growth-protostellar-cores}): (a) The
protostellar cores that form first are generally formed in the clumps
with the highest initial densities, and tend to have the highest final
masses. They accrete the bulk of their final mass from their close
vicinity. (b) On the other hand, matter that forms cores at later
times has already undergone considerable dynamical evolution; these
cores form from gas that was initially in widely distributed
low-density clumps. Along filaments, they stream towards a common
center of gravity and may merge at the intersections. Once enough mass
is accumulated, these clumps undergo rapid collapse and build up new
protostellar cores. The cores which form a late stages tend to have
very low final masses.

\item {\bf Competitive Accretion and the Importance of Dynamical
Interaction:} Once a gas clump becomes Jeans unstable it collapses and
forms a condensed core. This core grows in mass via accretion from the
infalling envelope. Merging may lead to clumps that contain multiple
cores. These compete with each other for the material of a common gas
reservoir. The succession of clump mergers leads to the formation of
an embedded, dense protostellar cluster, whose dynamical behavior is
dominated by close encounters between cluster members. Competitive
accretion and collisional dynamics determine the kinematical and
spatial properties of the cluster and the mass distribution of
protostellar cores (Sec.'s \ref{subsubsec:growth-protostellar-cores}
\& \ref{subsubsec:dyn-WW}).

\item {\bf Rotational Properties of Protostellar Cores:} The rotation
of a protostellar core is an important parameter for its collapse. It
determines the properties and stability of accretion disks and their
tendency for sub-fragmentation. Within the complex network of
intersecting filaments, gas clumps gain angular momentum from tidal
torque and shear. In the accretion process it is transfered onto the
embedded cores. Since the angular momentum is gained from large-scale
motion, the orientation of the spin vectors of individual protostellar
cores is correlated with their location
(Sec.~\ref{subsubsec:rotational-properties}). A similar correlation is
often found in observed star-forming regions between the orientation
of the molecular outflows from young stellar objects and their
location. However, the observational data are not conclusive.

\item {\bf Clustering Properties of Protostellar Cores:} The time
evolution of a highly Jeans-unstable region within a molecular cloud
leads to the formation of a dense cluster of protostellar cores. The
final dynamical state of the system closely resembles the properties
of observed stellar clusters: it exhibits the typical core/halo
structure of collision-dominated $N$-body systems, and when
calculating its 2-point correlation function, or equivalently the mean
surface density of companions as function of separation, one can
clearly distinguish between the binary regime and the large-scale
clustering regime (Sec.~\ref{subsubsec:clustering}).

\item {\bf Boundedness of Protostellar Clusters:} The clusters of
protostellar cores in our simulations form as bound entities: the
conversion of gas into condensed cores is such that the decrease of
potential energy is always balanced by the increase of kinetic energy
(Sec.~\ref{subsubsec:bound}).  Whereas the protostellar cluster is
bound, this may not be true for the resulting stellar cluster. Its
kinematical properties depend strongly on the details of the
conversion of individual cores into stars: on the speed and the
overall efficiency of the process.  However, this cannot be treated in
our simulations and needs to be addressed in detailed calculations of
individual core collapse.

\item {\bf Mass Spectrum of Protostellar Cores --- The Star Formation
Efficiency and Implications for the IMF:} The distribution of stellar
masses is one of the most important properties of the star-formation
process. Any comprehensive model of star formation must be able to
derive this quantity or at least address this issue. In our isothermal
models the masses of protostellar cores are the result of a sequence
of unpredictable stochastic events. In a natural way this leads to a
{\em log-normal} mass spectrum which peaks roughly at the {\em average
Jeans mass} of the system (Sec.~\ref{subsec:implications-IMF}).
Detailed collapse calculations show that perturbed rotating cores tend
to break up into multiple stellar systems, which cannot be resolved in
the larger-scale simulations presented here. Therefore, we compare the
numerical mass function of protostellar cores with the IMF derived for
multiple stellar systems. Our simulations require densities in the
range $n({\rm H}_2) = 10^5\,{\rm cm}^{-3}$ to $10^6\,{\rm cm}^{-3}$
for both distributions to agree.

\end{itemize}


\acknowledgements We thank Pavel Kroupa for valuable discussions on
the topic of binary stars and the IMF, Philippe Andr\'{e} for helpful
comments and remarks on the mass spectrum of prestellar condensations,
and Matthew Bate for his help with the SPH-code. We thank the referee
for many useful comments and for the prompt reply.

\begin{appendix}

\section[Clump Finding]{The Clump Finding Method}
\label{app:clumps}
This appendix describes the method used to identify the clumping
properties of the numerical calculations. It applies a scheme similar
to the one introduced by Williams et al.~(1994), but which is fully
integrated into the SPH formalism.

In SPH, the densities $\rho_i$ of individual particles $i$ are
obtained in a local averaging process over a list of neighbors within
distances less than twice the smoothing length $h_i$ (e.g.~Benz 1990,
Monaghan 1992). To identify clumps, the resulting 3-dimensional
density field is subdivided into ten bins equally spaced in the
logarithm of the density. Starting with the highest density level, the
particles are sorted by decreasing density, i.e.~the first particle in
the list is the one with the highest density. Going through this list,
for each particle $i$ it is checked whether the particles in its
neighbor list are already assigned to a clump. If this is not the case,
then particle $i$ is assigned to a new gas clump, together with all the
particles in its neighbor list. All additional particles at that
level which are connected to the particles of the new clump by means
of overlapping smoothing volumes are identified and assigned to the
same clump.  The assigned particles are finally removed from the
density list. On the other hand, if the neighbor list of the tested
particle $i$ contains contributions from already identified clumps the
particle will be associated to the clump that contributes the largest
fraction of particles in the neighbor list, only particle $i$ will be
removed from the sorted list. This scheme is repeated until all
particles at that density level are assigned to clumps. The procedure is
repeated at the next lower level and so on up to the lowest level.  
The method is
illustrated in Fig.~\ref{fig:clumps}. In the end, all SPH particles in
the system are assigned to individual clumps.

This scheme is free of assumptions about the geometrical shape of the
clumps\footnote{An alternative approach to determine clump properties
in (observed) molecular clouds was introduced by \cite{stu90}.  These
authors explicitly assumed that the gas clumps have Gaussian density profiles.
They decomposed their intensity maps into clumps with such
a profile by
minimizing the residual.} which is of great advantage when dealing with
highly irregular and filamentary structure as in our numerical
simulations of gravitational fragmentation and collapse.
Note that two clumps that may at a high level of
density be well separated entities, may at lower levels have common
contour lines. In our scheme, they are still separated even at these
lower levels by introducing an artificial `interface' between the two
clouds.  Individual particles are {\em  always} assigned to the clump
that contributes most neighbors. This sets a clear division line
between two competing clumps and enables their separation.  Besides
this basic separation criterion, the method is free of independent
parameters and uses only intrinsic properties of the SPH scheme.

\section{The Zel'dovich approximation and its validity to gaseous
systems}
\label{app:zel}
\subsection{The Zel'dovich Approximation}
In 1970, Zel'dovich proposed for cosmological simulations a method to
extrapolate the linear theory into the non-linear regime. The
dynamical evolution of an idealized self-gravitating and pressureless
continuous medium can be expressed in terms of a function ${\vec
f}({\vec r}_0,\delta t)$. Its value is the position $\vec r$ of a fluid
element after a time interval $\delta t$, whose original position was
${\vec r}_0$, i.e.~${\vec r} = {\vec f}({\vec r}_0,\delta t)$. If we
denote the original density field as $\rho({\vec r}_0)$ then the
density field after $\delta t$ is given by
\begin{equation}
\rho({\vec r},\delta t) = \frac{\rho({\vec r}_0)}{|{\rm det}
f_{i,j}({\vec r}_0,\delta t)|}\;.
\end{equation}
Here, $f_{i,j}$ denotes the partial derivative of the $i$-th component
of $\vec f$ in ${\vec r}_j$. The time evolution is given by Poisson's
equation and to first approximation it follows
\begin{equation}
\label{eqn:zel-shift}
{\vec r}(\delta t) = {\vec f}({\vec r}_0,\delta t) = {\vec r}_0 + {\vec v}({\vec r}_0)\,\delta t\;,
\end{equation}
with $\vec{\nabla}_{{\vec {\scriptsize r}}_{\scriptscriptstyle 0}}
\cdot {\vec v}({\vec r}_0) \propto \rho({\vec r}_0)$. This assumes
that the velocity field is rotation-free, because then the existence
of a potential $\phi\:\!({\vec r}_0)$ with ${\vec v}\;\!({\vec r}_0)
\equiv \vec{\nabla}_{{\scriptsize {\vec r}}_{\scriptscriptstyle 0}}
\phi({\vec r}_0)$ is guaranteed and is connected to the density field
via $\Delta_{{\scriptsize {\vec r}}_{\scriptscriptstyle 0}} \phi\:\!({\vec
r}_0) \propto \rho({\vec r}_0)$.

Using this method for the extrapolation of initially small density
fluctuations into the non-linear
regime (i.e.~into regions where the density contrast exceeds the value
of one) has certain limitations. For example, if $|{\rm det} f_{i,j}|$
becomes very small or zero, pressure forces would become
important, preventing infinite densities. Furthermore, if one follows
trajectories along which $|{\rm det} f_{i,j}|$ vanishes, the density
decreases again after the singularity and at the same time the
solution is no longer unique.  For practical purposes by
appropriately choosing the shift time $\delta t$ one can minimize these
problems and the Zel'dovich (1970) approximation is valid. We will
show that this also applies for gaseous systems. This procedure  considerably
speeds up the computation of the early evolutionary phase since the
system is advanced in one single large step as opposed to solving the
complete set of hydrodynamic equations and integrating over many small time
intervals.

We use the following method to generate random Gaussian fields:
After the desired power spectrum $P(k)$ is defined
we generate a {\em hypothetical} density field from all contributing
modes in Fourier space (according to Sec.~\ref{subsec:initial}). This
field is transformed back into real space and  Poisson's equation is
solved in order to obtain the velocities which would generate the
density field self-consistently. Starting from a homogeneous initial density
distribution, this velocity field is used to advance the particles in
the system in one single large time step. As illustration,
Fig.~\ref{fig:zel-shift} plots the 2-dimensional projection of the
homogeneous starting condition (Fig.~\ref{fig:zel-shift}a) and of the system after the
Zel'dovich shift has been applied for various shift intervals $\delta t$ 
(Fig.~\ref{fig:zel-shift}b -- h). 
Note that the method implicitly assumes periodic boundary
conditions. Therefore, the images have to be seen periodically
replicated in all directions.

\subsection[Validity of the Zel'dovich Approach]{Validity of the Zel'dovich Approach}
\label{subsec:validity}
To test the validity of the Zel'dovich approximation and its parameter
dependence we compare systems which have been generated with different
sets of shift intervals $\delta t$ and slopes ${\nu}$ of the power
spectrum.  Furthermore, we compare the results with systems that have
been advanced in time using the full SPH formalism.

If particles move along trajectories on which $|{\rm det} f_{i,j}|$
vanishes the result is no longer unique, i.e.~different paths may lead
to the same location. For converging flows, the density increases
before reaching the singularity in $|{\rm det} f_{i,j}|$ and decreases
afterwards.  This is unphysical for collision dominated
systems. Depending on the size and the strength of individual
fluctuations, the time interval $\delta t$ to reach this singular
point may differ. Assuming comparable amplitudes, for small
perturbations $\delta t$ will be short and for large-scale modes
$\delta t$ will be longer. This has to be taken into account when
determining the optimum choice of the shift interval. The effect of
varying $\delta t$ is addressed in Fig.~\ref{fig:zel-shift} which
shows a sequence of particle distributions generated by the Zel'dovich
method from a power spectrum $P(k) \propto k^{-2}$ with $0 \le \delta
t \le 30$. The homogeneous starting field is identical to a shift of
$\delta t=0$. Increasing $\delta t$, the density contrast starts to
grow and the distribution becomes more structured. However, at $\delta
t \approx 3$ small-scale fluctuations begin to disperse again and the
density contrast starts to diminish. At $\delta t\approx 30$ the
system appears homogeneous again. The same is true for different
slopes ${\nu}$ of the power spectrum. Figure~\ref{fig:zel-shift}i
illustrates the dependence of the average (open circles) and maximum
(solid circles) particle density on the time shift $\delta t$. The
dashed line denotes the density of the homogeneous cube as reference.

Figure~\ref{fig:corr-pow-dt} describes the influence of different
shift intervals $\delta t$ on (a) the 2-point correlation function
$\xi(r)$ and on (b) the measured power spectrum $P(k)$. In analogy to
the above, the correlation strength and length increase with shift
interval $\delta t$ for systems generated with $t\lesssim3$ and
decrease again for larger shifts. Similar conclusions apply to the
power spectrum. For $\delta t\approx2$ the initial slope of ${\nu}=-2$
is best reproduced. Shorter shifts $\delta t$ do not establish the
mode spectrum sufficiently, the spectrum is too flat. Longer intervals
produce overshooting on small scales, i.e.~perturbations with short
wave lengths are smeared out, whereas large modes are still
amplified. Therefore the spectrum gets steeper. For $\delta t>10$,
overshooting occurs for the largest modes as well and the entire power
spectrum flattens again.

Finally, we compare systems which have been generated by the
Zel'dovich method with systems that have been advanced in time solely
with SPH.  The Zel'dovich shift generates fluctuations on all scales,
since only gravitational forces are considered. Large perturbations
are Jeans unstable and start to collapse. Most small ones are Jeans
stable and pressure forces smear them out in the subsequent evolution
with the SPH method. However, also some low-mass fluctuations may have
been generated with sufficiently high density to be gravitationally
unstable, since the Jeans mass is inversely proportional to the square
root of the density.  These clumps will collapse as well. Furthermore,
while flowing towards a common center of gravity small clumps may
merge to form more massive clumps, which again may exceed the Jeans
limit. The probability for that to happen depends on the time scale
for the collapse and dispersal of individual clumps relative to the
time scale for clumps to merge or fragment. This can  be
determined only in a statistical sense.

If we advance the system from the homogeneous state with the full SPH
formalism, we compute the self-consistent initial velocity field of
the hypothetical fluctuation field.  Pressure forces are included from
the beginning and small perturbations have no possibility to grow
unless their mass exceeds the Jeans limit of the homogeneous system,
which is determined by the mean density. Therefore, only high mass
clumps can form. This is different from the Zel'dovich case, where
perturbations are created on all scales and also some low-mass clumps
may have large enough densities such that their local Jeans mass is
smaller than the clump mass. Therefore, slightly smaller and more
fragments are expected in systems that have been generated using the
Zel'dovich method. The detailed discrimination of two systems subtly
depends on the desired inhomogeneity and density contrast in the
system, i.e.~on the choice of ${\nu}$ and $\delta t$. These
small-scale differences decrease, the more homogeneous the
distribution and the smaller the Zel'dovich shift.  On large scales,
the statistical properties are much less affected.

This is exemplified, using a distribution of $50\,000$ particles with
a power spectrum $P(k) \propto k^{-2}$. Again the simulation cube
contains roughly 222 Jeans masses.  Figure~\ref{fig:comp-zel} shows
snapshots of the time evolution of the system initially generated
applying a Zel'dovich shift with $\delta t=2$ and subsequently
advanced with SPH. On the other hand, Fig.~\ref{fig:comp-sph}
describes the evolution of the system that was evolved from the
homogeneous state using SPH without applying the Zel'dovich method.
The large-scale behavior of the two systems is very similar, however
differences on small scales occur. The distribution initially evolved
with the Zel'dovich approximation is patchier at comparable
times. Note, that time is measured from the begin of the evolution
with SPH. To compare both systems at equal times, the Zel'dovich shift
interval $\delta t=2$ has to be added in
Fig.~\ref{fig:comp-zel}. However, with the progression of the SPH
calculation the higher degree of irregularity is reduced. Small
perturbations are smoothed by pressure.

In summary, the Zel'dovich approximation is very well suited to
generate fluctuation fields with well defined statistical properties
determined by the power spectrum $P(k)$. With the appropriate choice
of the shift interval $\delta t$, every spectrum can be
generated. When applying the Zel'dovich method to gaseous systems, one
has to take the effect of neglecting pressure forces into account. On
small scales, small deviations from the fully self-consistent time
evolution may occur.  Our test calculations do however
show that applying the Zel'dovich method is fully appropriate 
for the purpose of the current investigation.

\end{appendix}

\clearpage
\newpage



\clearpage
\newpage
{\bf\Large{Tables:}}\\

\begin{table}[p]
{\caption{Summary of the numerical models, all with temperature $\alpha =0.01$
and power spectrum $P(k) \propto1/k^2$.}\label{tab:models-N=2}}
\vspace{-0.25cm}
\begin{center}
\begin{tabular}[t]{cccl}
\hline
{{Model}} & {{Particle}} & {{Initial}\hfill} & {{Zel'dovich}} \\
& {{Number}}\hfill & {{Distribution}}\tablenotemark{\it a}  &
{{Shift}}\tablenotemark{\it b}\hfill  \\
\hline
 $\cal A$ & $50\;\!000$  & random & $\delta t = 2.0$ \\
 $\cal B$ & $50\;\!000$  & random & $\delta t = 2.0$ \\
 $\cal C$ & $50\;\!000$  & random & $\delta t = 2.0$ \\
 $\cal D$ & $50\;\!000$  & random & $\delta t = 2.0$ \\
 $\cal E$ & $50\;\!000$  & random & $\delta t = 1.0$ \\
 $\cal F$ & $50\;\!000$  & grid   & $\delta t = 2.0$ \\
 $\cal G$ & $200\;\!000$ & random & $\delta t = 2.0$ \\
 $\cal H$ & $200\;\!000$ & random & $\delta t = 1.5$ \\
 $\cal I$ & $500\;\!000$ & random & $\delta t = 1.5$ \\
\hline
\end{tabular}
\end{center}
{\tablenotetext{\it a}{Initial distribution for the Zel'dovich
shift. A  homogeneous random distribution is denoted by `random',
whereas `grid'  means that the
particles have initially been placed on a regular grid.}}
{\tablenotetext{\it b}{Shift interval $\delta t$ for the Zel'dovich approach.}}
\end{table}

\begin{table}[p]
{\caption{\label{tab:moments}Moments of the density and of the line-of-sight velocity
centroid pdf }}
\begin{center}
\begin{tabular}[t]{crrrr}
\hline
 &{$x\;\;$} & {$y\;\;$} & {$z\;\;$} & {$\log_{10}\rho$} \\
\hline
\multicolumn{5}{l}{\rule{0cm}{0.0cm}}\\[-0.1cm]
\hline
\multicolumn{5}{l}{$t=0.0$ --- no cores have formed yet}\\
\hline
$\mu$     & $-0.03$ & $-0.02$ & $-0.02$ & $-0.69$\\ 
$\sigma$  &  $0.02$ &  $0.03$ &  $0.02$ &  $0.38$\\ 
$\theta$  &  $0.06$ & $-0.04$ & $-0.09$ &  $0.22$\\ 
$\kappa$  &  $1.00$ &  $1.00$ &  $1.01$ &  $2.91$\\ 

\multicolumn{5}{l}{\rule{0cm}{0.0cm}}\\[-0.4cm]
\hline
\multicolumn{5}{l}{$t=1.2$ --- 6 cores with $M_{\ast}=1\%$}\\
\hline
$\mu$     & $-0.03$ & $-0.02$ & $-0.01$ & $-0.68$\\ 
$\sigma$  &  $0.04$ &  $0.07$ &  $0.12$ &  $0.51$\\ 
$\theta$  & $-0.34$ & $-0.16$ & $-0.34$ &  $1.52$\\ 
$\kappa$  &  $5.11$ &  $4.30$ &  $2.19$ &  $7.19$\\ 

\multicolumn{5}{l}{\rule{0cm}{0.0cm}}\\[-0.4cm]
\hline
\multicolumn{5}{l}{$t=1.5$ --- 15 cores with $M_{\ast}=10\%$}\\
\hline
$\mu$     & $-0.03$ & $-0.01$ & $-0.01$ & $-0.56$\\ 
$\sigma$  &  $0.07$ &  $0.13$ &  $0.21$ &  $0.78$\\ 
$\theta$  & $-0.20$ & $-0.49$ & $-0.07$ &  $1.47$\\ 
$\kappa$  &  $6.59$ &  $3.53$ &  $1.95$ &  $5.30$\\ 

\multicolumn{5}{l}{\rule{0cm}{0.0cm}}\\[-0.4cm]
\hline
\multicolumn{5}{l}{$t=2.0$ --- 31 cores with $M_{\ast}=30\%$}\\
\hline
$\mu$     & $-0.04$ &  $0.00$ & $-0.07$ & $-0.42$\\ 
$\sigma$  &  $0.11$ &  $0.20$ &  $0.30$ &  $0.98$\\ 
$\theta$  &  $0.91$ & $-0.34$ &  $0.20$ &  $0.84$\\ 
$\kappa$  &  $7.31$ &  $4.48$ &  $2.64$ &  $3.04$\\ 

\multicolumn{5}{l}{\rule{0cm}{0.0cm}}\\[-0.4cm]
\hline
\multicolumn{5}{l}{$t=2.8$ --- 51 cores with $M_{\ast}=60\%$}\\
\hline
$\mu$     & $-0.07$ &  $0.03$ & $-0.06$ & $-0.28$\\ 
$\sigma$  &  $0.19$ &  $0.36$ &  $0.37$ &  $1.14$\\ 
$\theta$  &  $0.55$ & $-0.15$ &  $0.03$ &  $0.28$\\ 
$\kappa$  &  $6.31$ &  $4.83$ &  $4.92$ &  $2.61$\\ 

\multicolumn{5}{l}{\rule{0cm}{0.0cm}}\\[-0.4cm]
\hline
\multicolumn{5}{l}{$t=3.9$ --- 56 cores with $M_{\ast}=90\%$}\\
\hline
$\mu$     & $ 0.00$ &  $0.08$ & $-0.06$ & $-0.44$\\ 
$\sigma$  &  $0.32$ &  $0.36$ &  $0.34$ &  $1.17$\\ 
$\theta$  & $-0.70$ &  $0.77$ &  $1.61$ &  $0.42$\\ 
$\kappa$  &  $8.59$ &  $7.55$ & $11.73$ &  $3.47$\\ 
\hline

\end{tabular}
\end{center}
The first four moments $\mu$, $\sigma$, $\theta$, and
$\kappa$, respectively, as defined in Eqn.'s \ref{eqn:mom1} to
\ref{eqn:mom4}, of the distribution of the line-of-sight velocity
centroids ($x$-, $y$-, and $z$-component) and of the logarithm of the
density $\log_{10}\rho$ for the model $\cal I$ at six different stages
of the dynamical evolution.
\end{table}

\begin{table}[ht]
\caption{\label{tab:fitting} Optimum fit parameters}
\vspace{0.2cm}
\renewcommand{\arraystretch}{1.5}
\begin{center}
\begin{tabular}[p]{cccccc}
\hline
& $\log_{10} \,\mu$   & $\log_{10} \,\mu$  & $\log_{10} \,\sigma$   & $n({\rm H}_2)$\\
& ($\mu$ in $\langle M_{\rm J}\rangle$)  & ($\mu$ in  ${\rm M}_{\odot}$) &
($\sigma$ in ${\rm M}_{\odot}$)  & (in cm$^{-3}$)\\
\hline
 $M_*=30\;\!$\% & $\!\!\!\!-0.06$   & $-0.31$ & $-0.53$ & $2.3 \times
10^5$ \\
 $M_*=60\;\!$\% & $0.15$            & $-0.31$ & $-0.43$ & $6.2 \times
10^5$ \\
 $M_*=90\;\!$\% & $0.31$            & $-0.31$ & $-0.49$ & $1.3 \times
10^6$ \\
 IMF   & --- & $-0.31$ & $-0.38$  & --- \\
\hline
\end{tabular}
\end{center}

 {Parameters of the optimum log-normal fits to the combined core mass
distribution in Fig.~\ref{fig:mass-spectrum-3}. The lines $M_*=30$\%,
$M_*=60$\%, and $M_*=90$\% denote the evolutionary stages where 30\%,
60\%, and 90\% of the available gas mass is accreted onto protostellar
cores.  The entry IMF indicates the parameters of  the mass function
of multiple stellar systems (Kroupa et al.\ 1990, model MS). The second 
column denotes the peak location of the log-normal fit function
(Eqn.\ \ref{eqn:log-normal-function}) with respect to the average Jeans mass
in the system. Column 3 and 4, show  peak and width in
units of solar mass for the physical scaling where core mass spectrum
and IMF peak at the same position (i.e.\ when all entries in column 3
agree).  The fifth column lists the corresponding physical average
gas density in the system. }
\end{table}
\renewcommand{\arraystretch}{1.0} 

\clearpage
\newpage
{\bf \Large Figures:}\\

\begin{figure}[p]
\caption[F01]{\label{fig:nine-models}  Initial particle distribution for the
nine SPH simulations with $\alpha=0.01$ and $P(k) \propto 1/k^2$ projected
into the $xy$-plane.  The six models with $50\;\!000$ particles: a)
$\cal A$, b) $\cal B$, c) $\cal C$, d) $\cal D$, e) $\cal E$, f) $\cal F$.
Models with $200\;\!000$ particles: g) $\cal G$, h) $\cal H$. The
high-resolution model with $500\;\!000$ particles: i) $\cal I$. Every
figure contains $50\;\!000$ particles. In case of a larger total
number, the plotted particles are selected randomly. Model $\cal F$ (f)
is generated from particles placed on a grid, which is still visible
in the figure. All other models use random uniform particle
distributions. The Zel'dovich shift applied to generate model $\cal E$
is $\delta t=1.0$, the shift for models $\cal H$ and $\cal I$ is $\delta
t=1.5$. Otherwise, $\delta t=2.0$ is used. Therefore these
distributions exhibit more structure and larger density contrasts.
For a summary see Tab.~\ref{tab:models-N=2}.}
\end{figure}

\begin{figure}[p]
\caption[F02]{\label{fig:3D-cube-A} Snapshots of the model $\cal A$ at
$t=0.0$ (initial condition at the start of the SPH simulation), at
$t=0.2$, at $t=0.5$, at $t=0.7$, at $t=1.0$, and at $t=1.2$. During
the dynamical evolution, small-scale fluctuations are damped, whereas
Jeans unstable clumps start to collapse.  At $t=0.5$, the first
compact cores (`sink particles' in terms of the SPH code) have formed
in the centers of the densest clumps and have accreted $M_*=5$\% of the
total gas mass. At $t=0.7$ the protostellar cores altogether gained
10\% of the available mass. Each of the following snapshots of the
system is taken at the phase where the mass of the cluster of cores
has increased further by 10\% of the total mass. Gas particles
are plotted as small dots and collapsed cores are denoted by thick
dots. Note that the figures do not give information about the
smoothing volume of individual particles. Its size is such that it
contains roughly 50 neighbors. }
\vspace{0.25cm}
\end{figure}

\begin{figure}[p]
{Figure~\ref{fig:3D-cube-A} --- continued:} Snapshots of the model
$\cal A$ at $t=1.4$, $t=1.6$, $t=1.8$, $t=2.1$, $t=2.5$, and
$t=3.3$. The total gas mass accreted onto protostellar cores is
$M_*=40$\%, $M_*=50$\%, $M_*=60$\%, $M_*=70$\%, $M_*=80$\%, and
$M_*=90$\%, respectively.
\end{figure}

\begin{figure}[p]
\caption[F03]{\label{fig:3D-cube-I} Density distribution of the
high-resolution model $\cal I$ at four different times: Initially, at
$t=1.5$, at $t=2.0$, and at $t=2.8$. The total mass fraction accreted
onto protostellar cores is denoted by $M_*$. The location of
protostellar cores is denoted by black dots and the density is scaled
logarithmically.}
\end{figure}

\begin{figure}[p]
\caption[F04]{\label{fig:rho-pdfs}  Density pdf for
the high-resolution simulation $\cal I$ at six different stages of the
dynamical evolution of the system. (a) The initial state at
$t=0.0$. (b) At $t=1.2$ when the first few cores have formed and
contain about 1\% of the total mass. (c) At $t=1.5$ when protostellar
cores account for $M_*=10\%$ of the system mass. (d) At $t=2.0$ and
$M_*=30\%$, (e) at $t=2.8$ and $M_*=60\%$, and finally (f) at $t=2.9$
where $M_*=90\%$, i.e.~the cluster of cores contains almost all mass
in the system. The dashed lines in some plots describe an linear fit
with the slope $d\log_{10}N/d\log_{10}\rho$ indicated by the attached
number.  }
\end{figure}

\begin{figure}[p]
\caption[F05]{\label{fig:vel-pdfs}  Line centroid
velocity pdf for the high-resolution simulation $\cal I$ at four
relevant stages of the dynamical evolution for the line-of-sight
aligned with the $x$-, $y$-, and $z$-axes of the system: (a) At
$t=1.2$ when the first few cores have formed and contain about
$M_*=1\%$ of the total mass. (b) At $t=2.0$ when is $M_*=30\%$. (c) At
$t=2.8$ and $M_*=60\%$, and finally (d) at $t=2.9$ where almost all
gas mass is accreted onto cores, $M_*=90\%$. The dashed parabola in
the main plots indicate the Gaussian distribution obtained from using
the first two moments $\mu$ and $\sigma$ from
Tab.~\ref{tab:moments}. For neither pdf those give a good fit. The
inlay in the upper right corner of each figure indicate the total line
profile ($v$ versus $\log_{10}N$) obtained from summing over the
velocity contribution of all particles in the system.  }
\end{figure}

\begin{figure}[p]
\caption[F06]{\label{fig:clump-spectrum-Z} Properties of identified clumps
and condensed cores in the high-resolution simulation $\cal I$ at
times $t=0.0$, 1.5, 2.0 and 2.8, i.e.~when 0\%, 10\%, 30\% and 60\% of
the total gas mass is collapsed onto protostellar cores. For each
point in time the upper panel shows the distribution of gas clump
masses (thin lines) and core masses (thick lines). The lower panel
shows the location of each identified clump in a density--mass
diagram. A detailed explanation is given in the main text. }
\end{figure}

\begin{figure}[p]
\caption[F07]{\label{fig:accretion-history}   Formation and accretion history
of selected protostellar cores in the simulations (a) $\cal A$ and (b)
$\cal I$. The numbers reflect the order of their formation.  }
\end{figure}

\begin{figure}[p]
\caption[F08]{\label{fig:sink-masses} Final masses $m$ of the protostellar
cores $i$ in the high-resolution model $\cal I$ (black dots) and in
model $\cal A$ (stars) for comparison. Again the numbers reflect the
order of their formation.}
\end{figure}

\begin{figure}[p]
\caption[F09]{\label{fig:mass-region}   Initial distance distribution of
matter that gets accreted onto protostellar cores. For (a) the first
nine cores to form and for (b) the last nine ones, the number $N$ of SPH
particles that get accreted onto a condensed core are plotted as
function of their initial distance $r$ to the particle that is converted
into the accreting core particle. The cores that form first accrete
more mass than the later ones and this mass is accumulated from a
smaller volume.}
\end{figure}

\begin{figure}[p]
\caption[F10]{\label{fig:clump-accretion} Mass contribution $m_j$ of
individual clumps $j$ identified in the initial density field to the
final mass of a core particle $i$: for (a) core $i=1$, (b) core $i=2$,
(c) core $i=10$, (d) core $i=40$ and (e) core $i=55$. The total mass
of the gas clumps $j$ is indicated by dots, their mass contribution to
core $i$ by crosses. Plot (f) gives the mass-weighted mean of the
fraction of clump material $\left \langle f_i \right \rangle$ that
gets accreted onto individual cores $i$. Cores that form early (small
values of $i$) tend to accrete nearby clumps more completely, whereas
material that builds up cores at later stages has already undergone
sufficient dynamical evolution to be well mixed. The initial
association with certain clumps becomes irrelevant.}
\end{figure}

\begin{figure}[p]
\caption[F11]{\label{fig:trajectory}   a)
Trajectories of cores \#3, \#5, \#11, \#31 and \#46, projected into
the $xy$-plane. b) Magnification of the unstable triple (\#3,\#5,\#11)
that forms early at the beginning. Core \#5 gets expelled at $t\approx
0.8$.  c) Enlarged view on the trajectories during the second
major dynamical event at $t \approx 1.6$: the hardening of binary
(\#3,\#11) in an encounter with object \#46 and subsequent acceleration
due to interaction with particle \#31.}
\end{figure}

\begin{figure}[p]
\caption[F12]{\label{fig:ind-clumps}   
 Typical shapes of identified clumps in the high-resolution model
$\cal I$ at time $t=1.2$, i.e.~when four collapsed cores have formed
and accreted 1\% of the total gas mass. This first part of the figure
projects the clumps containing the four protostellar cores into the
$xy$-, $xz$- and $yz$-plane. Density is scaled logarithmically. Two
contour lines span one decade ($\log_{10}
\Delta \rho = 1/2$) with the lowest contour level being  $10^{1/2}$ above the
mean density $\left \langle \rho \right \rangle = 1/8$. The black dots
indicate the positions of the dense protostellar cores. }
\end{figure}

\begin{figure}[p]
{Figure~\ref{fig:ind-clumps} --- continued:}  Typical
shapes of identified clumps in the high-resolution model $\cal I$ at time
$t=1.2$. This second part of the figure shows contour plots of clumps
that have not yet collapsed to high enough densities to contain dense
cores. Note that clumps are numbered accoring to their peak
density. The plot is centered on the center-of-mass of each clump, as
indicated by the two intersecting lines.
\end{figure}

\begin{figure}[p]
\caption[F13]{\label{fig:radial-density-profile}Radial surface density
profiles for the $xz$-projection of clumps \#4 and \#12 in the
sample. These clumps are selected because their projections appear
relatively round and smooth (cf.~with
Fig.~\ref{fig:ind-clumps}). Hence, the radially averaged density
profile is a good approximation of the real density distribution. For
comparison, the projected surface density $\Sigma$ of an isothermal
sphere is indicated by the dotted line. It has a 3-dimensional density
profile $\rho(r) \propto (r^2+r_{\rm c}^2)^{-1}$ where $r_{\rm c} =
0.005$ is the core radius (which is equal to the radius of the sink
particle in model $\cal I$). Whereas the density profile flattens out
for clump \#12 which has yet to form a condensed core, the density for
clump \#4 increases all the way inwards to the central dense core. The
steep density fall-off at $\log_{10} R
{\:\lower0.6ex\hbox{$\stackrel{\textstyle >}{\sim}$}\:} -0.8$ is an
artifact of the clump search algorithm. At these radii the clumps
blend into the background gas and an artificial boundary is
introduced.}
\end{figure}

\begin{figure}[p]
\caption[F14]{\label{fig:larson-X}   Clustering properties of the
protostellar cluster. The left panel plots the location of protostellar
cores  projected into the $xy$-, the $xz$- and $yz$-plane,
respectively, at (a) $t=1.0$, (b) $t=2.0$,  (c)
$t=3.0$, and (d) $t=5.0$. The right panel shows the projected mean surface
density of companions $\xi$ for protostellar cores as function of
separation $r$.  Data are taken from model $\cal A$ since it has been
advanced furthest in time. }
\end{figure}

\begin{figure}[p]
\caption[F15]{\label{fig:binary-separations} Distribution of semi-major
axes for identified bound pairs of protostellar cores in model $\cal A$ at
times (a) $t=1.0$, (b) $t=2.0$, (c) $t=3.0$, and (d) $t=5.0$. The
times are chosen in analogy to the previous figure. For cores
associated with higher-order systems this gives only a rough estimate of
the orbital parameters.}
\end{figure}

\begin{figure}[p]
\caption[F16]{\label{fig:ang-momentum-1} Time evolution of the angular
momenta $\vec{L}$ of protostellar cores. The left column plots (a) the
$x$-component $L_x$, (c) the $y$-component $L_y$ and (e) the
$z$-component $L_z$ of each condensed core in the high-resolution
simulation $\cal I$ as function of time $t$. The right column plots
the distribution of every component at the end of the SPH
simulation. For model $\cal I$ at $t=3.9$ (solid lines) and for
comparison form for model $\cal A$ at $t=5.6$ (dashed lines): (b)
$L_x$, (d) $L_y$ and (f) $L_z$.}
\end{figure}

\clearpage
\begin{figure}[p]
\caption[F17]{\label{fig:ang-momentum-2} (a) Distribution of the total
angular momenta $|\vec{L}|$ of protostellar cores in the simulation
$\cal{I}$ at $t=2.8$ (solid lines), i.e.~when 60\% of the gas has been
accreted, and at the final stage $t=3.9$ (dashed lines). (b)
Distribution of the specific angular momenta $|\vec{L}|/m$ at the
same times. (c) Specific angular momentum $|\vec{L}|/m$ as function of
the mass $m$ of the cores (filled circles denote $t=2.8$ and open
circles the final state $t=3.9$).}
\end{figure}

\begin{figure}[p] 
\caption[F18]{\label{fig:ang-momentum-3} Projection of the angular momenta
$\vec{L}$ of protostellar cores in simulation $\cal{I}$ into the $xy$-, $xz$-
and $yz$-plane at $t=1.5$ (upper panel), $t=2.0$ (middle panel) and
$t=2.8$ (lower panel), i.e.~when 10\%, 30\% and 60\%, respectively, of
the gas mass is accreted onto the cores. The length of the lines is
proportional to $|\vec{L}|$; note that for better legibility the
scaling factor in the bottom row is reduced by a factor of 1/3
compared to the upper two panels.}
\end{figure}

\begin{figure}[p]
\caption[F19]{\label{fig:boundedness} Energetic properties of the
protostellar clusters that form in model $\cal A$ (dashed lines) and
model $\cal I$ (solid lines): Time evolution of (a) the total kinetic
energy, subdivided into the contribution from random motions $E_{\rm
int}$ (thick lines) and from the center-of-mass motion $E_{\rm cm}$
(thin lines at the bottom of the plot), (b) the potential energy
$E_{\rm pot}$, (c) the virial coefficient $\eta_{\rm vir} \equiv
2E_{\rm int}/|E_{\rm pot}|$, and (d) the cumulative mass $M_{\rm tot}$
of the cluster. }
\end{figure}

\begin{figure}[p]
\caption[F20]{\label{fig:velocity-dispersion}   Time evolution of the total
(3-dimensional) velocity dispersion $\sigma$ (thick lines), of the
line-of-sight velocity dispersion along the $x$-, $y$- and $z$-axis
(thin lines) and of the center-of-mass velocity (thin lines at low
velocities) of the protostellar cluster that forms during the
dynamical evolution of the standard model $\cal A$ (dashed curves) and
the high-resolution model $\cal I$ (solid curves).  }
\end{figure}

\begin{figure}[p]
\caption[F21]{\label{fig:mass-spectrum-1} Mass distribution of protostellar
cores in the high-resolution model $\cal I$ at different stages of the
dynamical evolution. In the upper right corner of each plot, the time
$t$ and the fraction of the total gas mass converted into protostellar
cores $M_*$ is indicated. Analog to Fig.~\ref{fig:clump-spectrum-Z},
the vertical line indicates the mass resolution limit of the simulation.}
\end{figure}

\begin{figure}[p]
\caption[F22]{\label{fig:mass-spectrum-2}  Distribution of protostellar core
masses for all models described in this chapter
(Tab.~\ref{tab:models-N=2}) at the stage when roughly 60\% of the gas
mass is accumulated in condensed cores: (a) corresponds to model
$\cal A$, (b) to model $\cal B$, and so forth. }
\end{figure}

\begin{figure}[p]
\caption[F23]{\label{fig:mass-spectrum-3} Combined distribution of
core masses in all simulations with $\alpha=0.01$ and
$P(k)\propto1/k^2$ at times when 30\%, 60\% and 90\% of the total gas
mass has been converted into protostellar cores. The open circles
denote the best fit Gaussian (see Tab.~\ref{tab:fitting}) and the
dashed curve in each plot specifies the IMF for multiple stellar
systems (Kroupa et al.\ 1990). The system is scaled such that the
maxima of both distributions agree --- recall that the core mass
distribution peaks roughly at the average Jeans mass of the
system. For comparison the dotted curve shows the original Miller \&
Scalo (1979) mass function (for constant star formation over $12\times
10^9$ years), shifted to the same peak mass. }
\end{figure}


\begin{figure}[p]
\caption[F25]{\label{fig:clumps} Illustration of the clump finding
algorithm in a 1-dimensional sample case: At the highest density level
one clump is identified and all particles at this level get assigned
to it. At the third level an additional clump is detected. At the
sixth highest level, a third new clump is identified. Clumps \#1 and
\#2 now have overlapping contour lines and are separated as described
above. This is indicated by an arrow. At level eight the last clump is
detected and separated from the other at the lowest level (again
indicated by arrows).  }
\end{figure}

\begin{figure}[p]
\caption[F26]{\label{fig:zel-shift} Projections of the particle
distribution generated by the Zel'dovich method from a power spectrum
$P(k) \propto k^{-2}$. (a) Homogeneous starting distribution.  The
distribution after applying the Zel'dovich method with (b) $\delta
t=1/2$, (c) $\delta t=1$, (d) $\delta t=2$, (e) $\delta t=3$, (f)
$\delta t=5$, (g) $\delta t=10$, and (h) $\delta t=30$. (i) The
dependence of the average and maximum particle density on the time
shift $\delta t$ is plotted in open and filled circles,
respectively. The dashed line denotes the average density of the
homogeneous cube.}
\end{figure}

\begin{figure}[p]
\caption[F27]{\label{fig:corr-pow-dt}  (a) 2-point
correlation function $\xi(r)$ and (b) power spectrum $P(k)$ for
different shift intervals $\delta t$. To compute the functions, the cube has
been subdivided into $(128)^3$ cells. The random homogeneous starting
distribution is plotted with dotted lines. The statistical errors for
different distances $r$ and wave numbers $k$ apply to all fields
equally.}
\end{figure}

\begin{figure}[p]
\caption[F28]{\label{fig:comp-zel} 
Snapshots of the evolution of a system generated by the Zel'dovich
method from a power spectrum $P(k) \propto k^{-2}$ and shift interval
$\delta t=2$. After applying the Zel'dovich shift, the subsequent evolution is
calculated using SPH. The figures show projections of
the 3-dimensional distribution into the $xy$-plane in intervals of
$\Delta t = 0.3$: (a) initially at $t=0.0$, (b) at $t=0.3$, (c) at
$t=0.6$, (d) at $t=0.9$, (e) at $t=1.2$, and (f) at $t=1.5$. Time is
measured from the begin of the SPH simulation.}
\end{figure}

\begin{figure}[t]
\caption[F29]{\label{fig:comp-sph}  
Analogous to Fig.~\ref{fig:comp-zel}, but describing a system that is
entirely evolved in time using SPH. No Zel'dovich shift is applied.
The system is projected into the $xy$-plane at (a) $t=0.0$, showing
the homogeneous starting condition, at (b) $t=1.0$, (c) $t=1.7$,
(d)$t=2.0$, (e) $t=2.3$, (f) at $t=2.6$, (g) at $t=2.9$, (h) at
$t=3.2$, and (i) at $t=3.5$. Again, time is measured from the begin of
the SPH simulation.}
\end{figure}

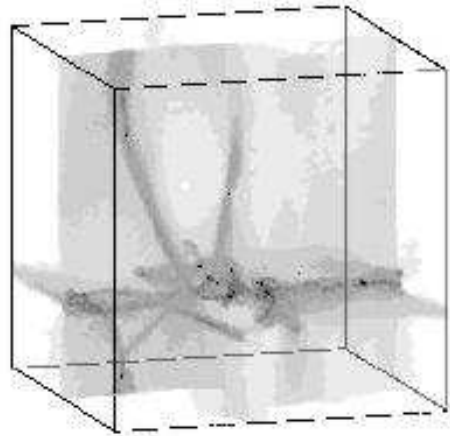
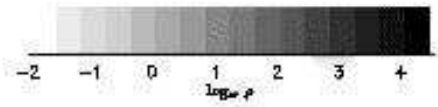
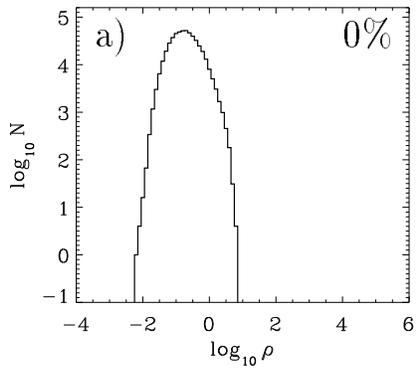
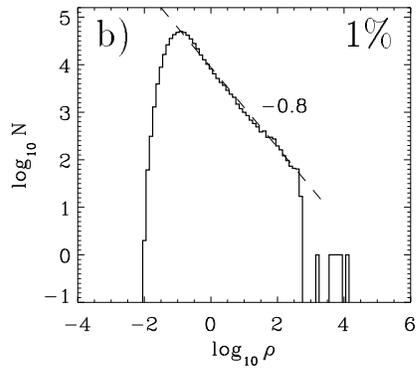
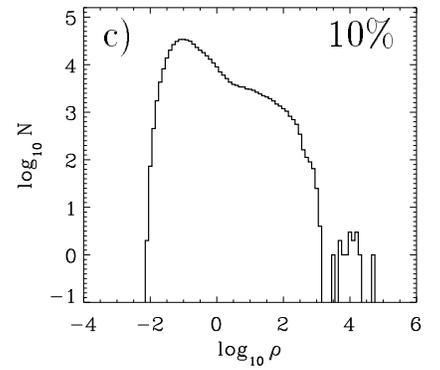
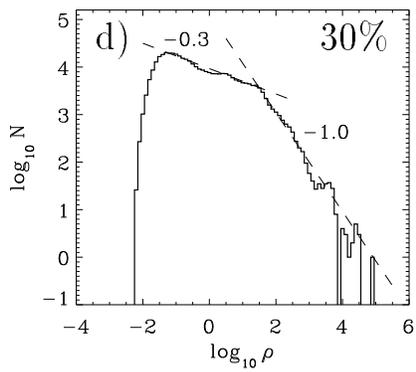
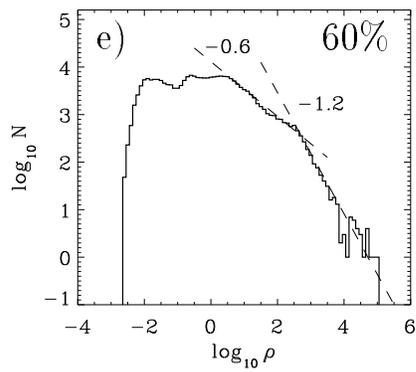
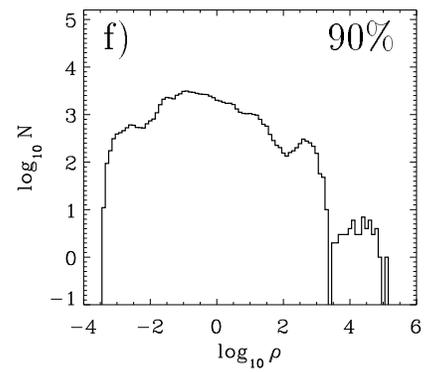
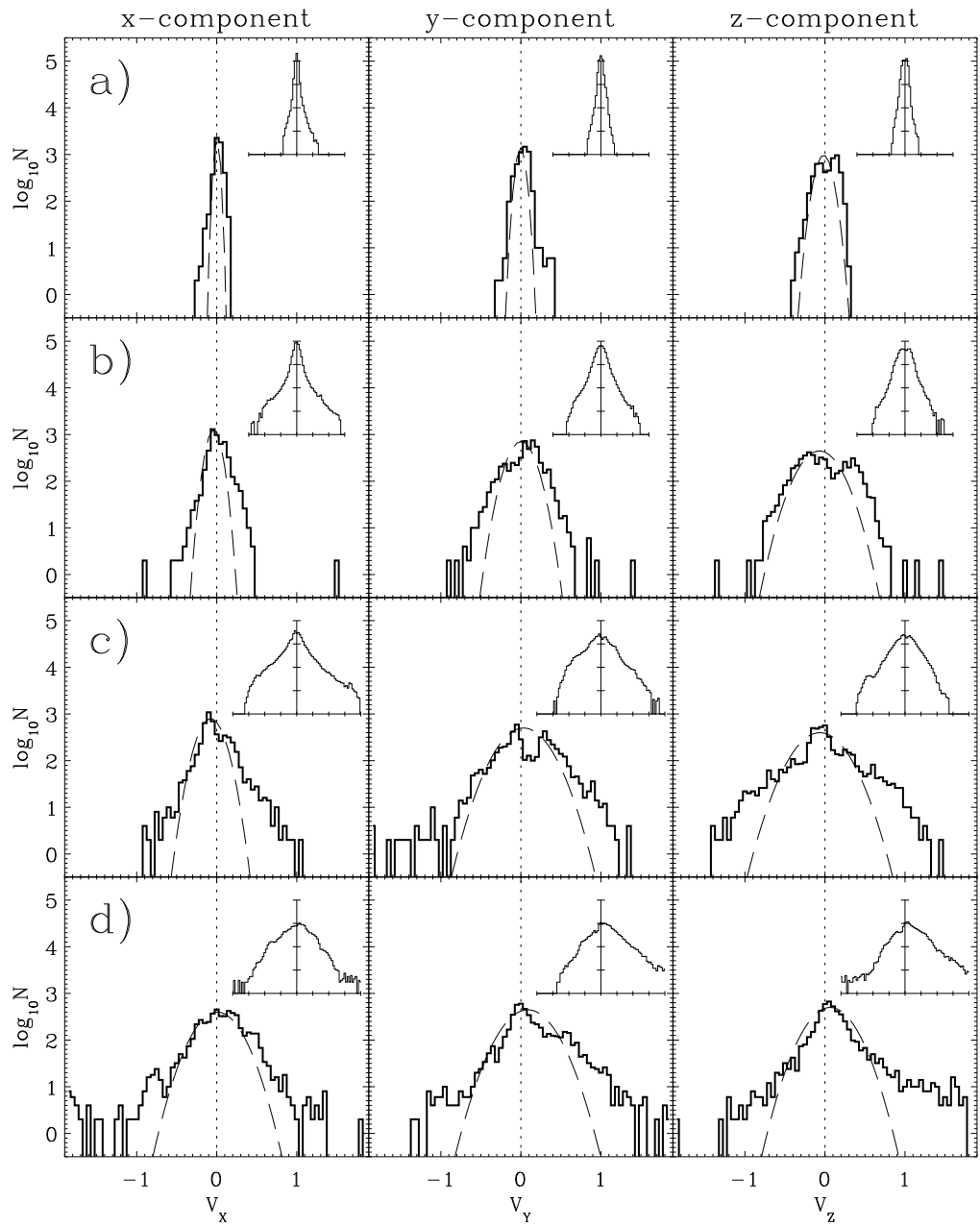
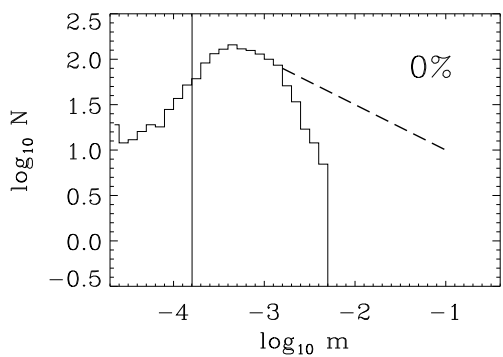
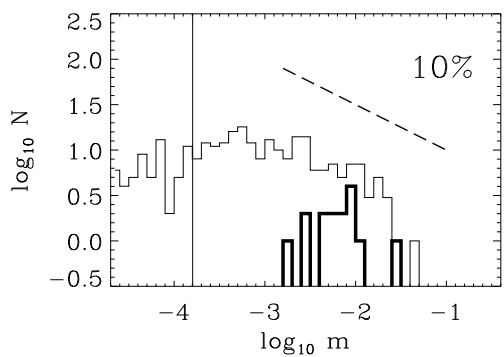
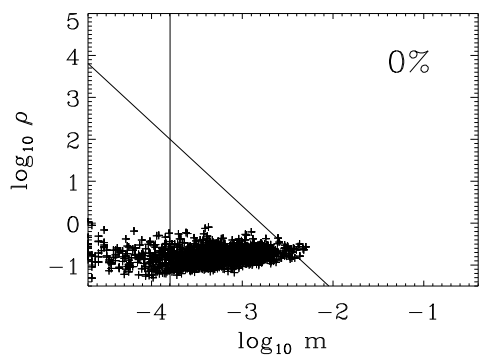
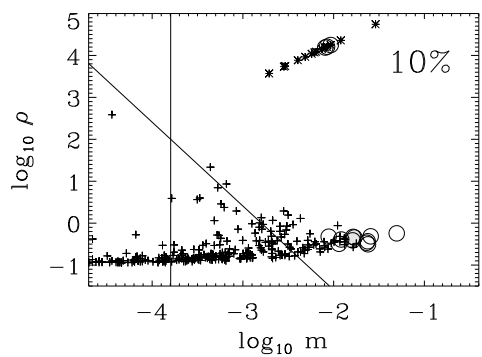
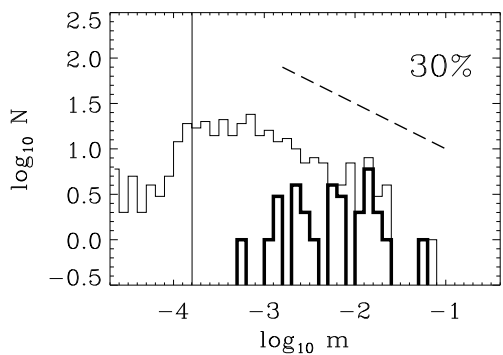
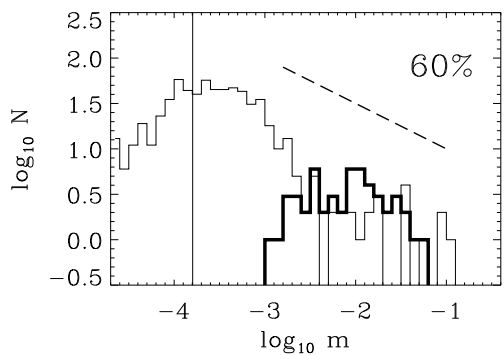
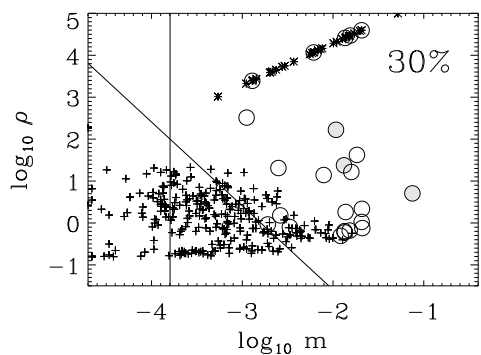
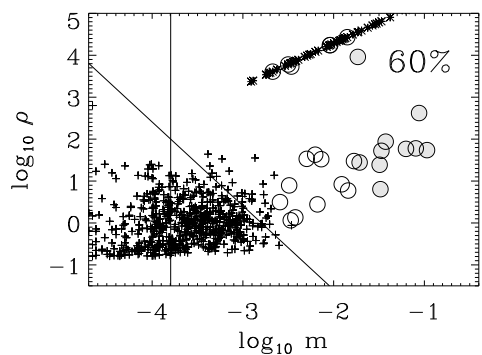
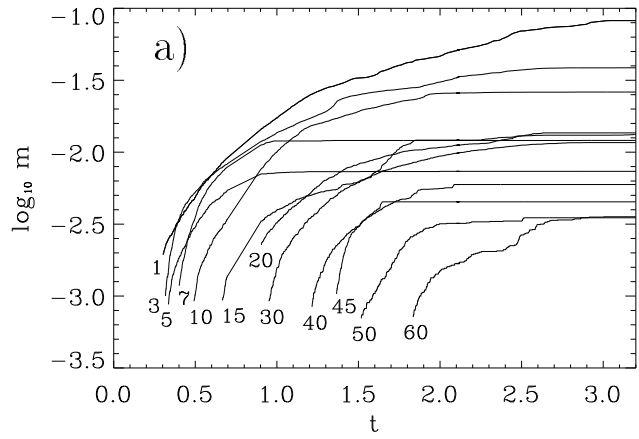
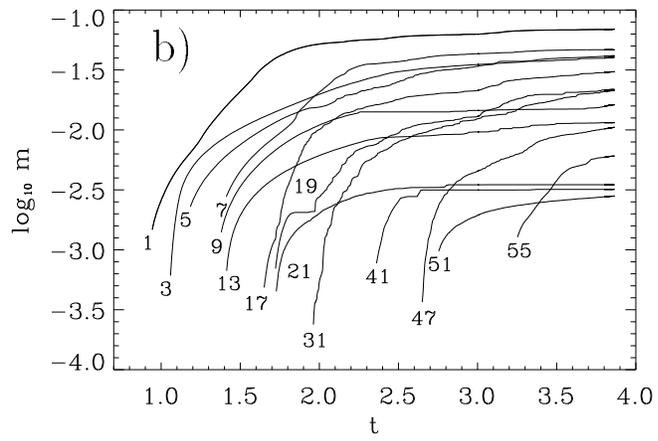
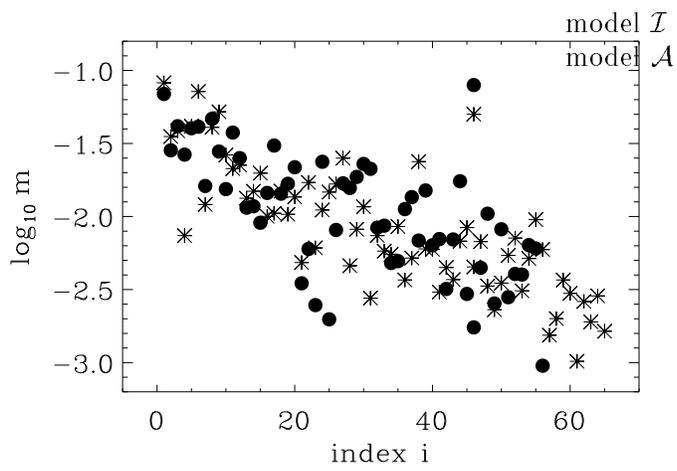
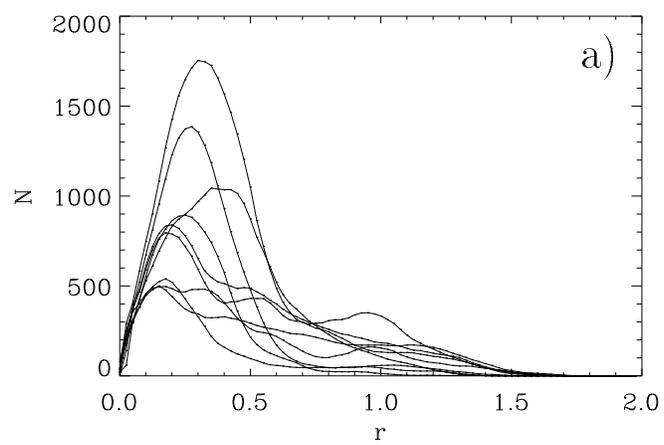
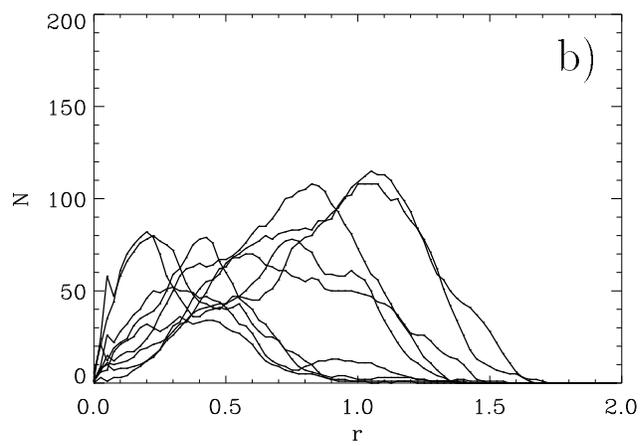
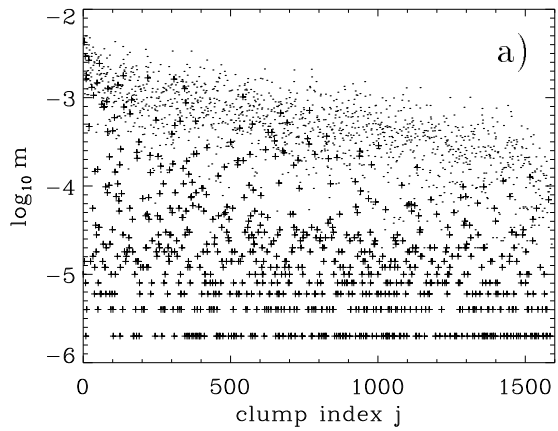
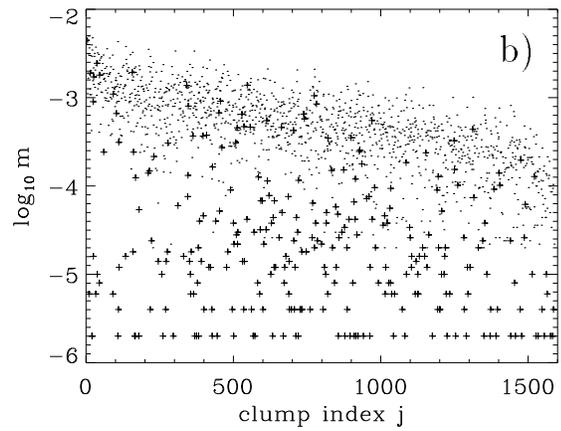
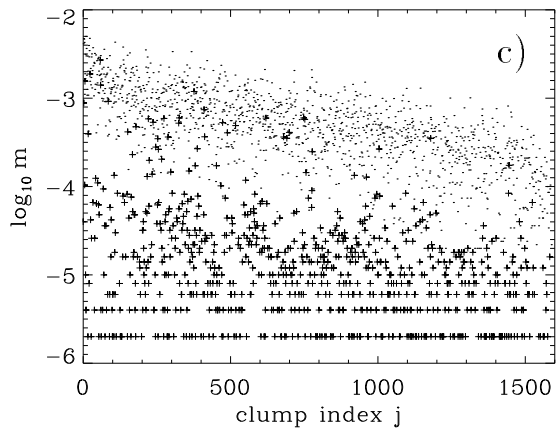
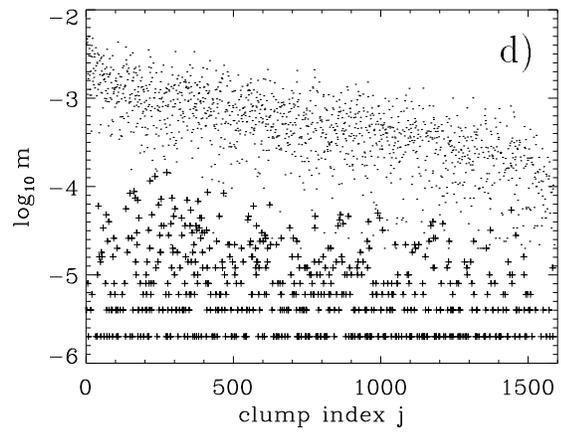
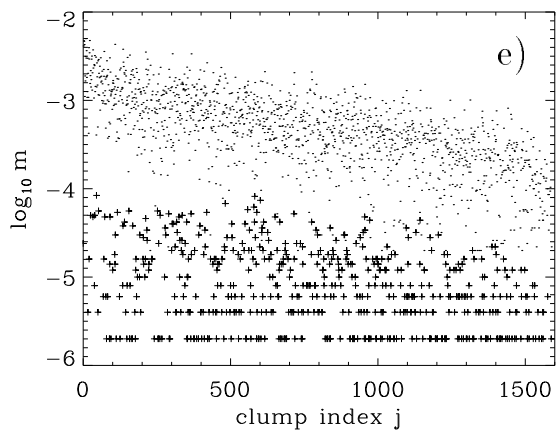
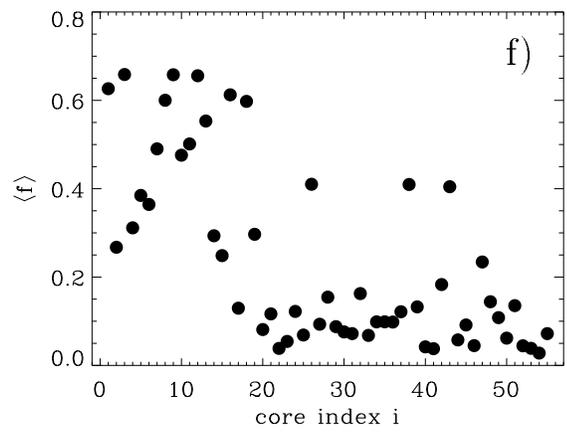
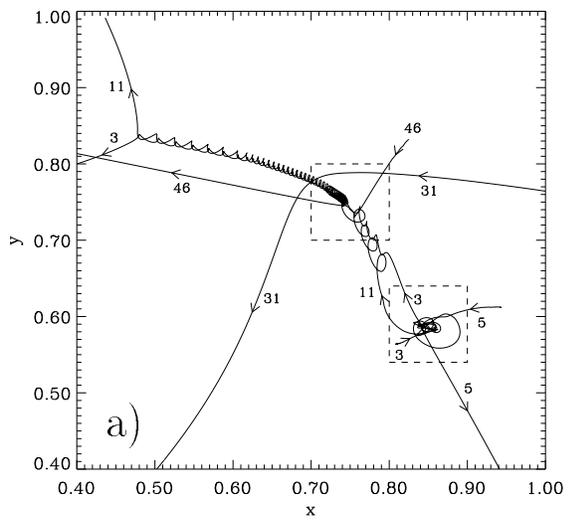
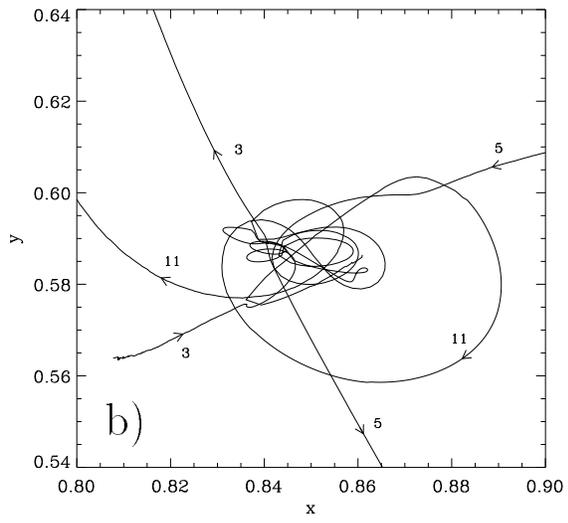
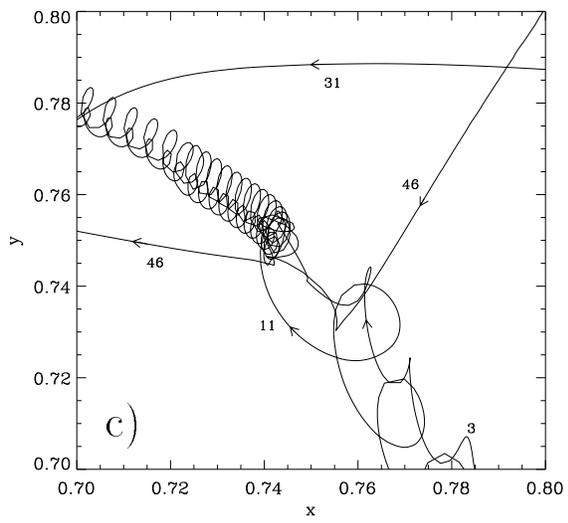
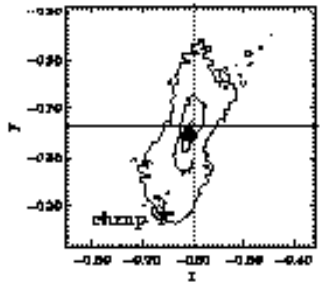
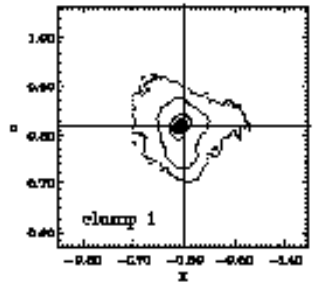
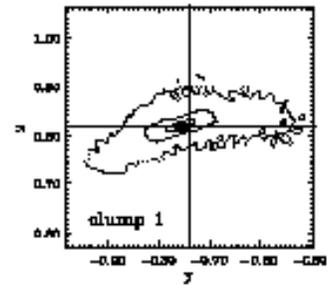
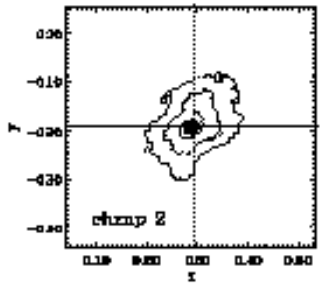
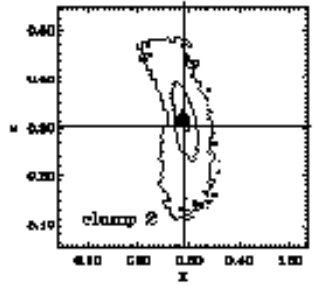
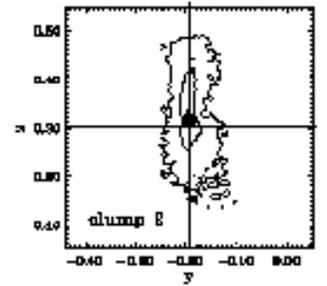
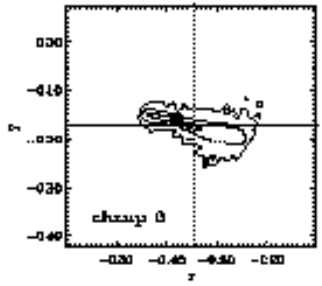
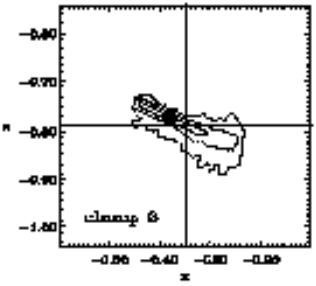
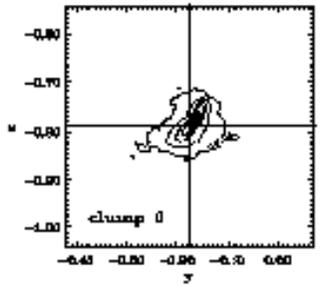
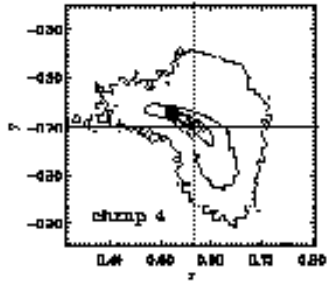
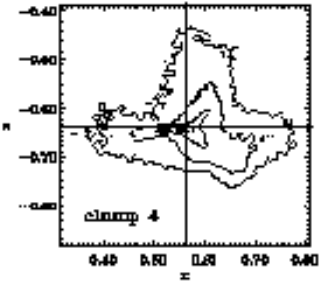
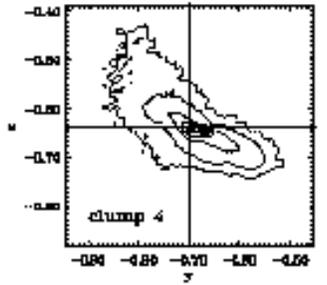
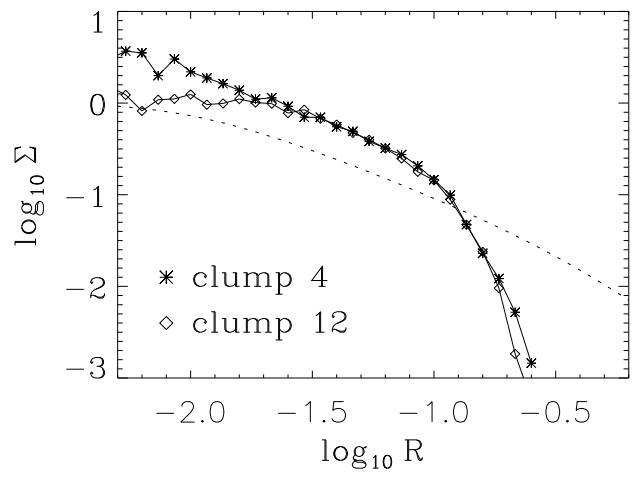
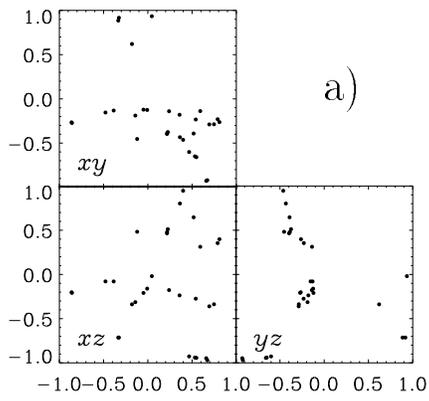
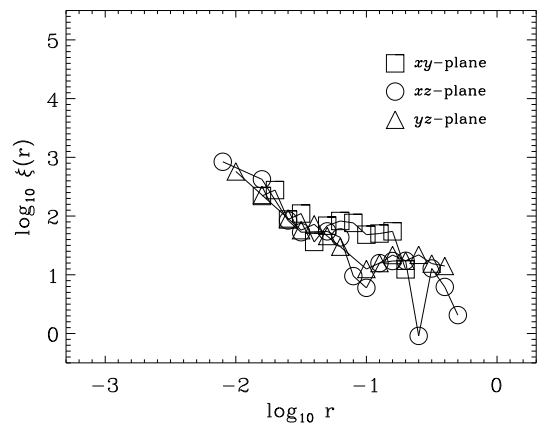
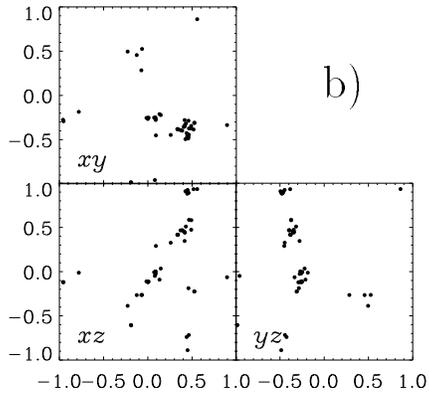
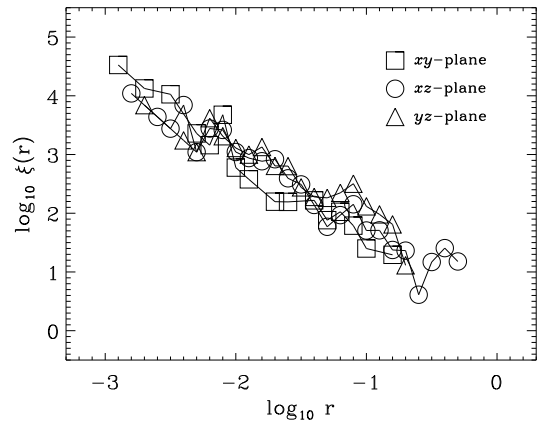
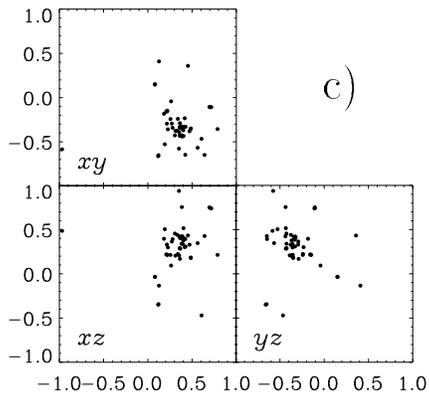
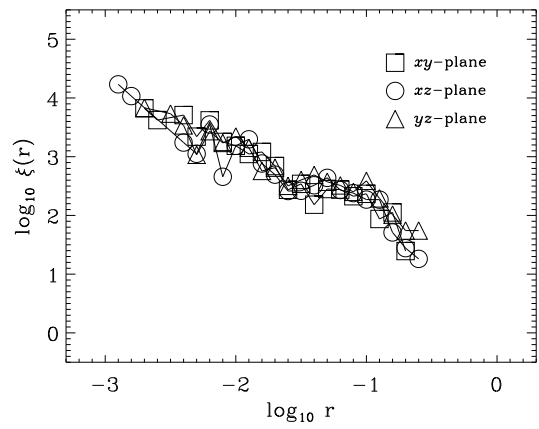
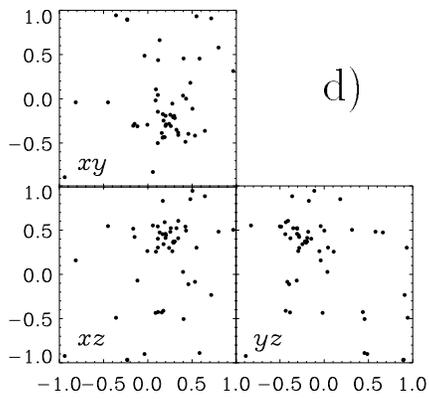
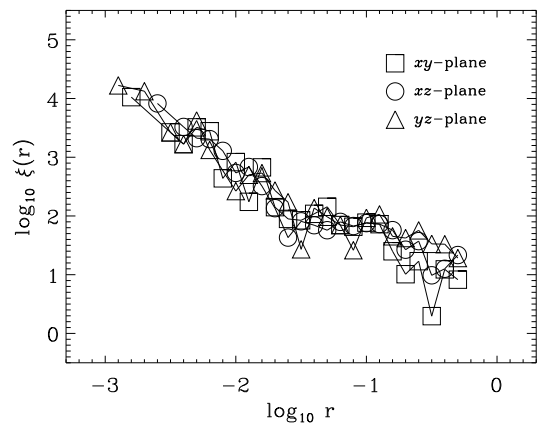
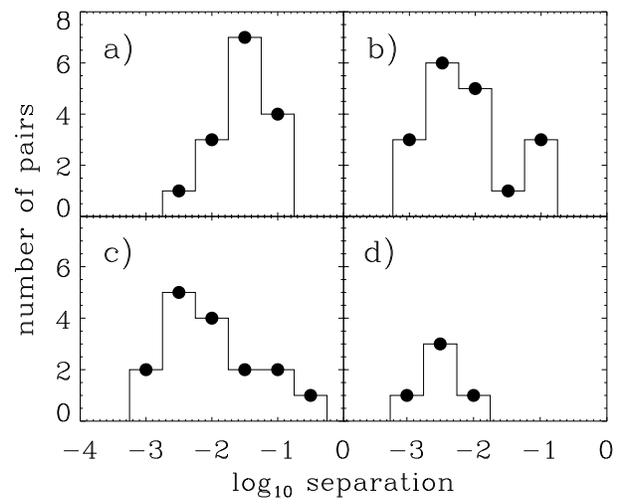
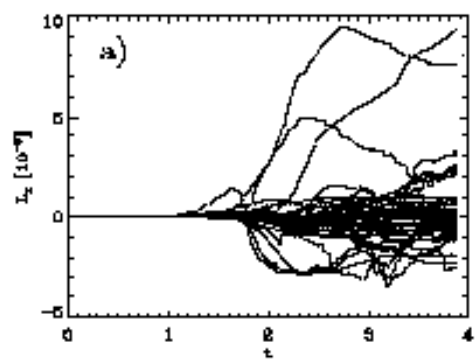
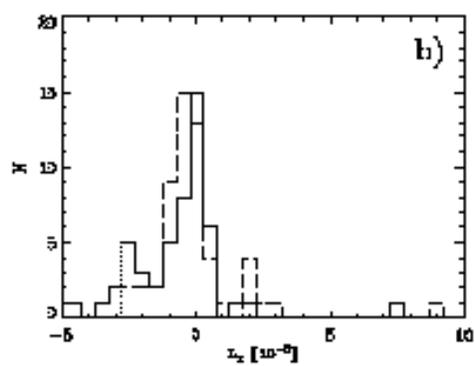
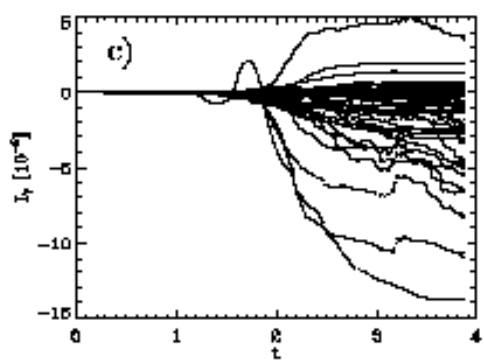
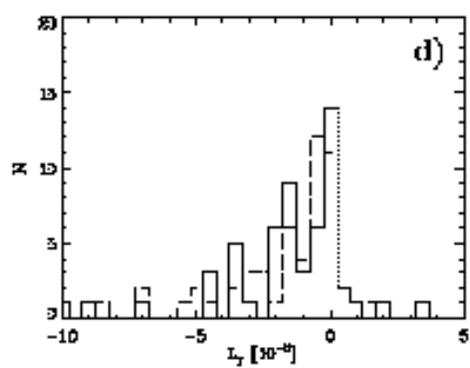
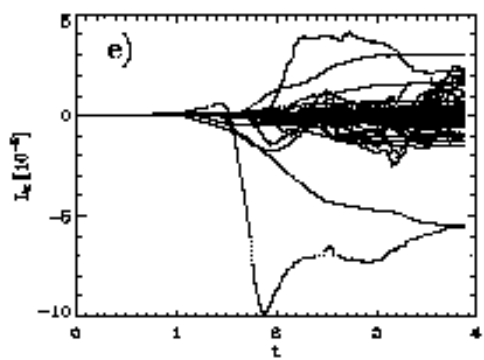
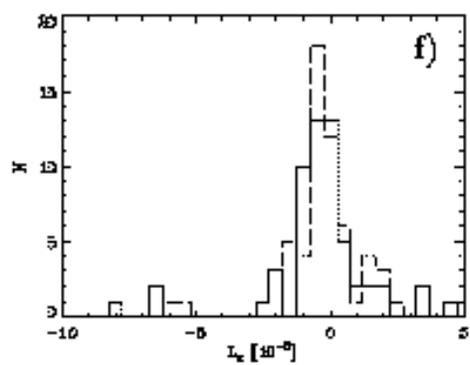
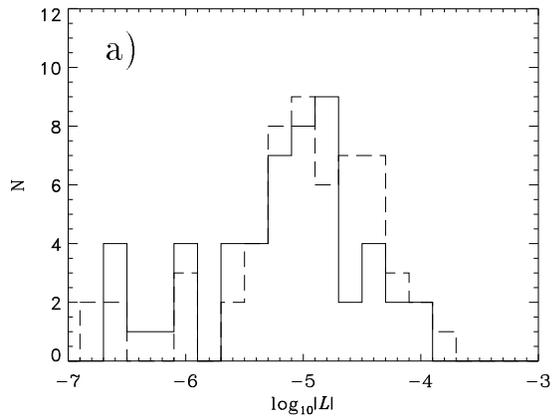
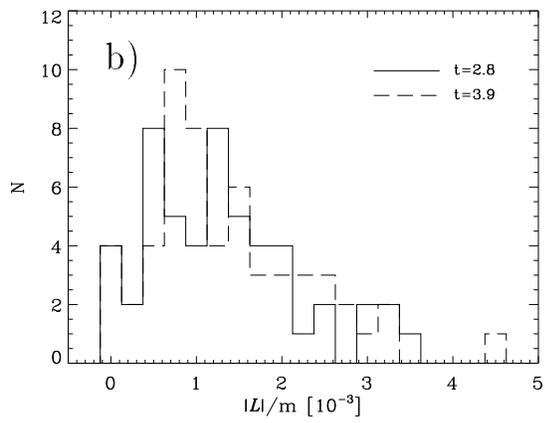
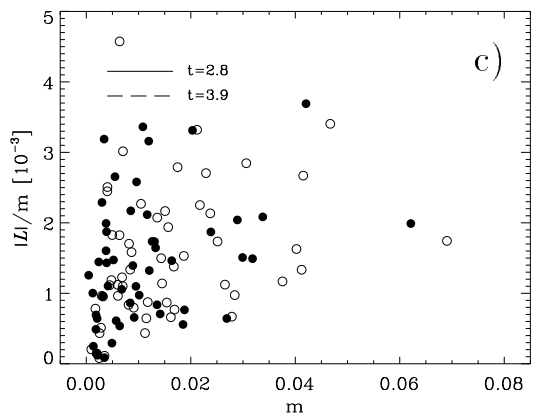
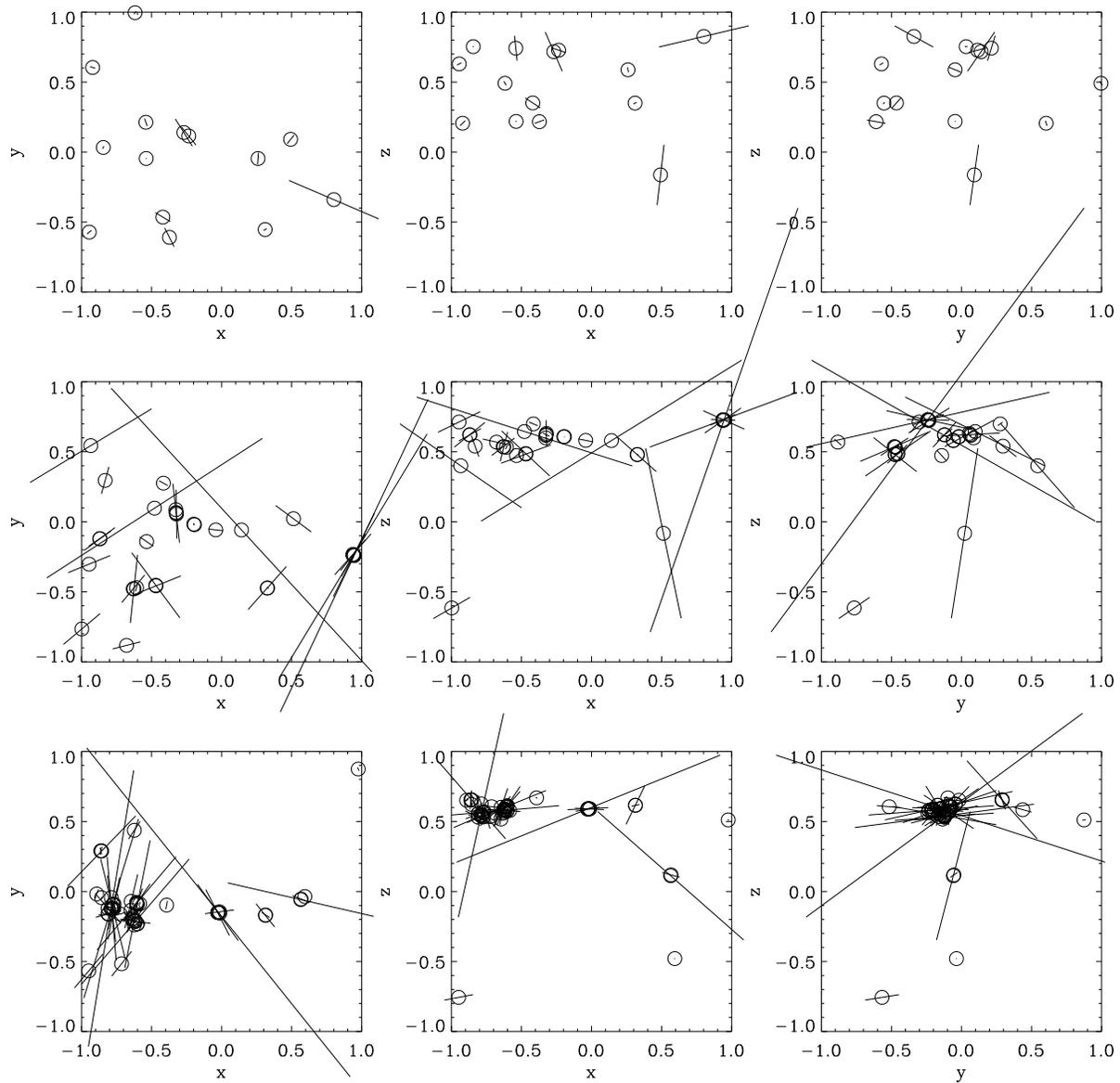
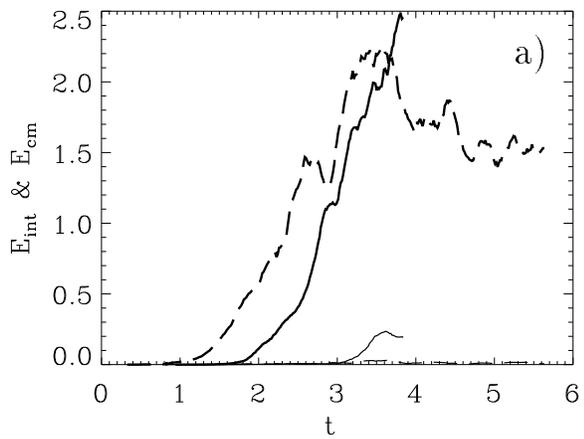
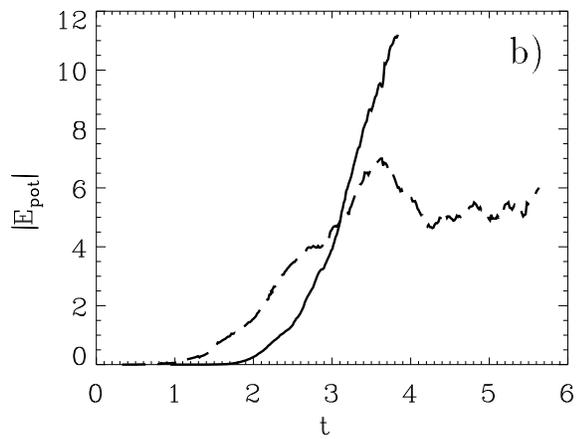
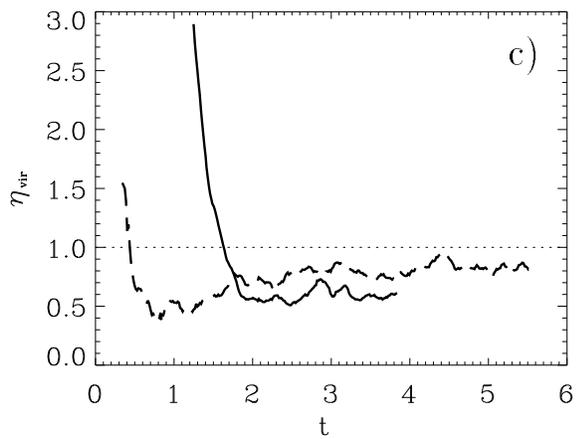
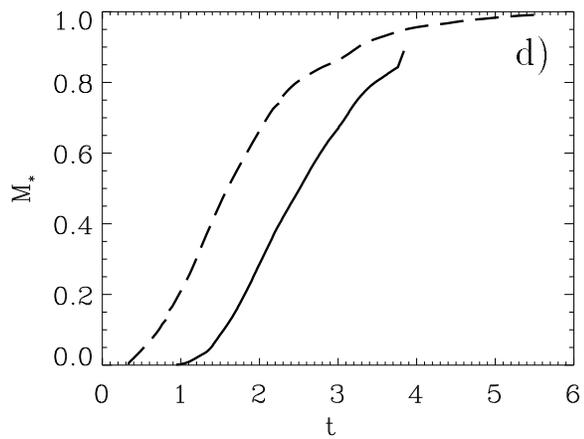
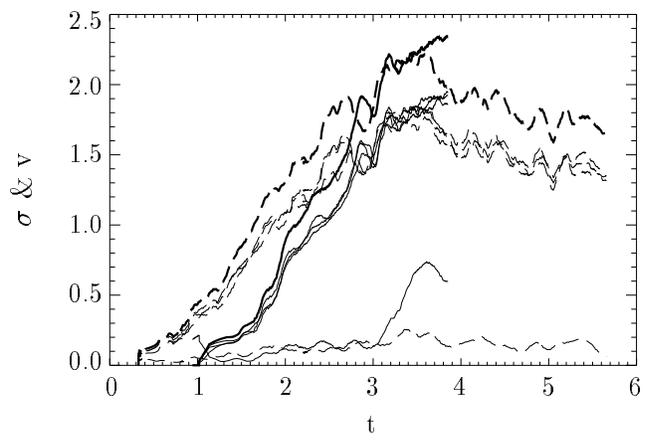
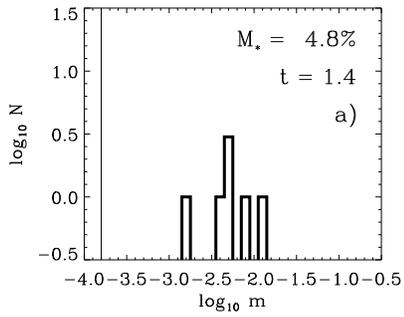
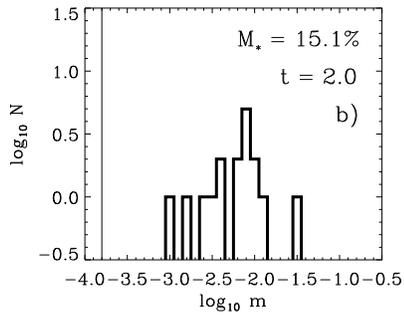
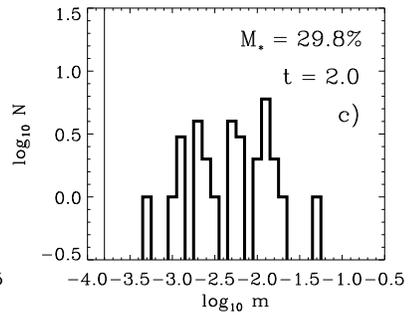
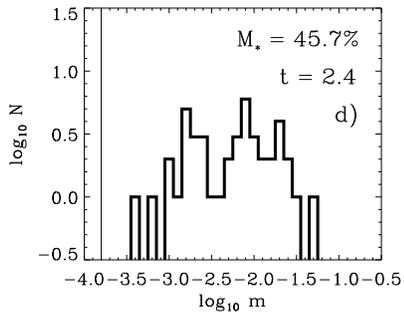
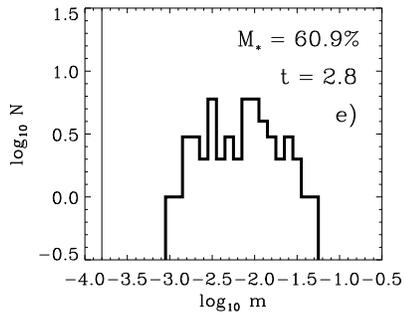
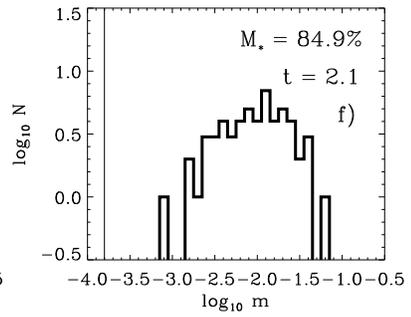
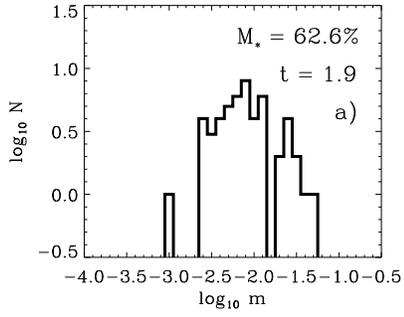
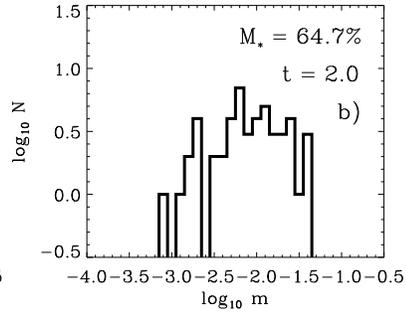
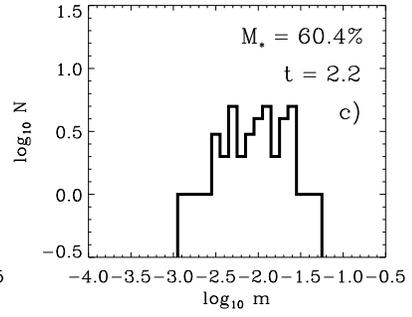
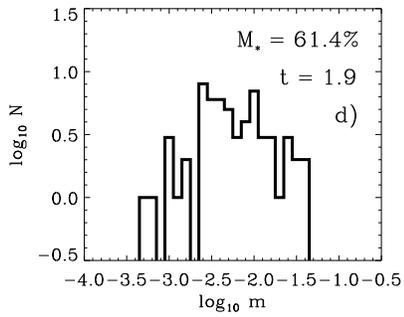
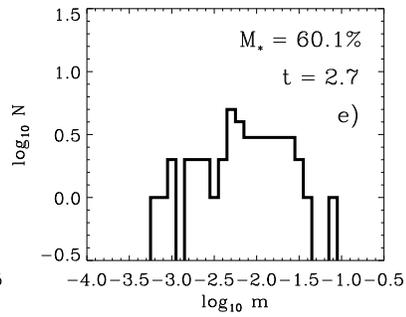
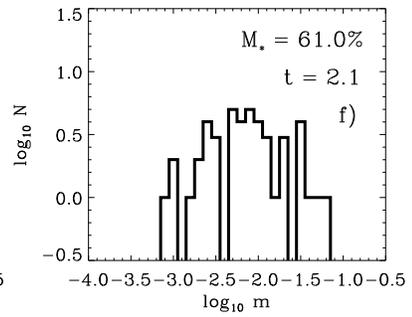
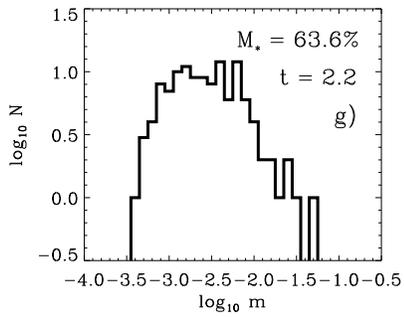
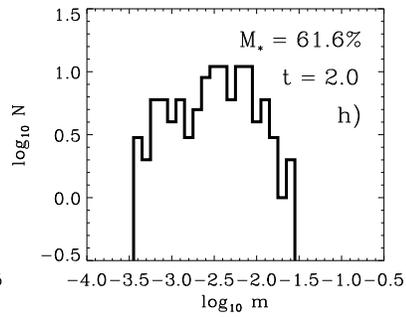
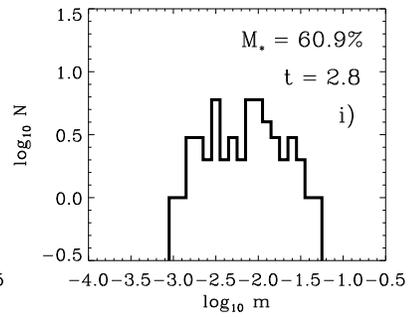
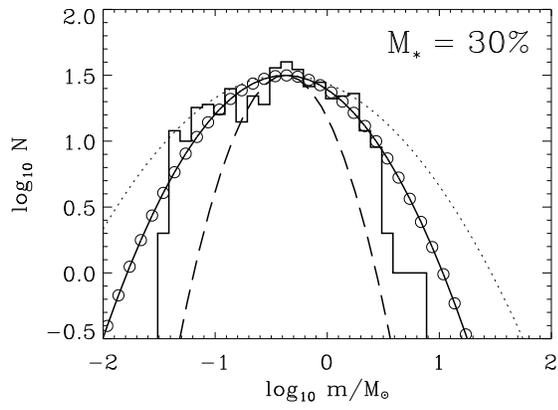
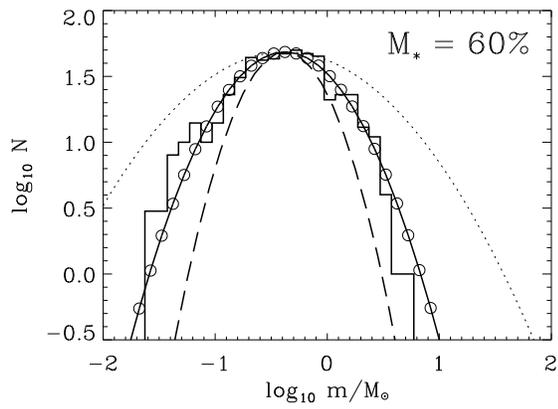
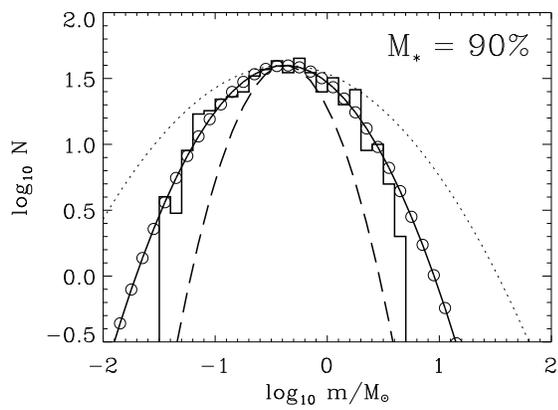
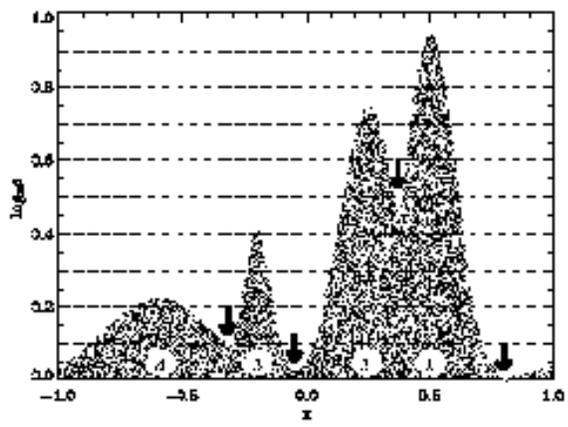
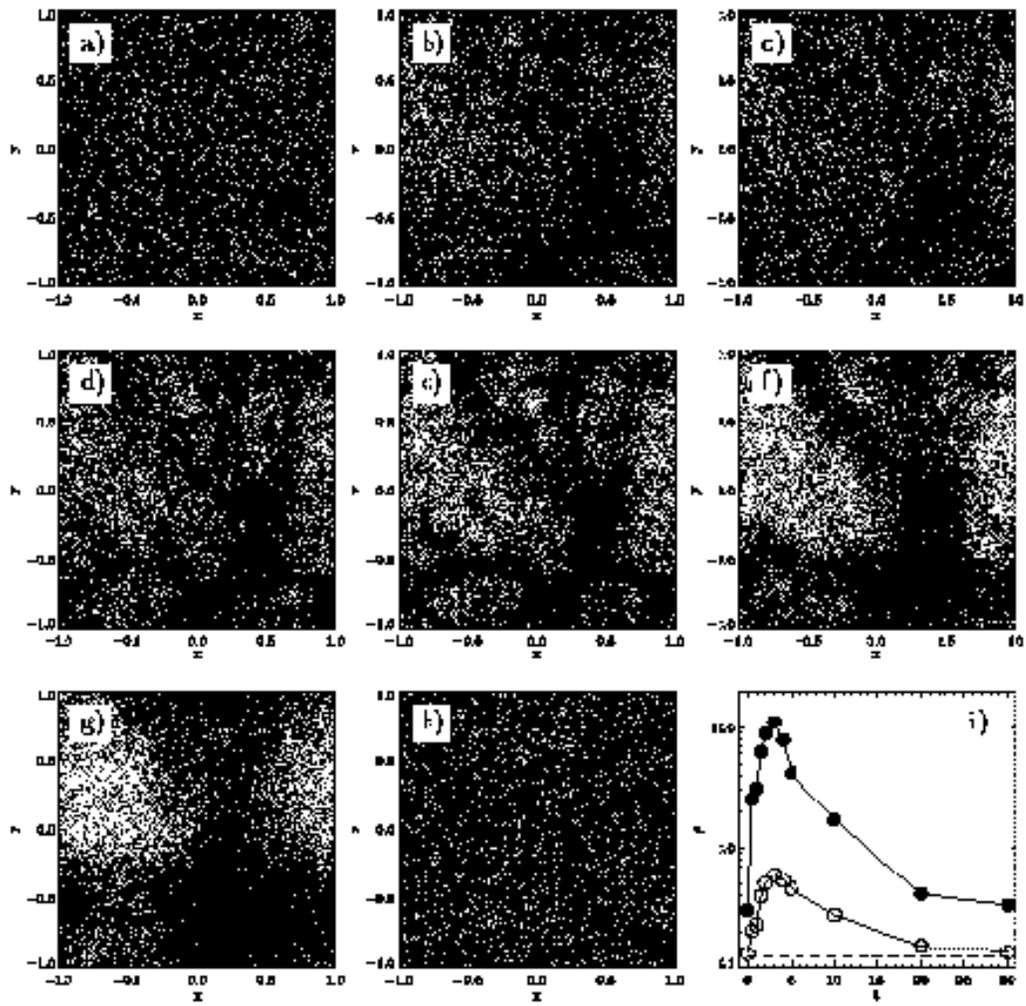
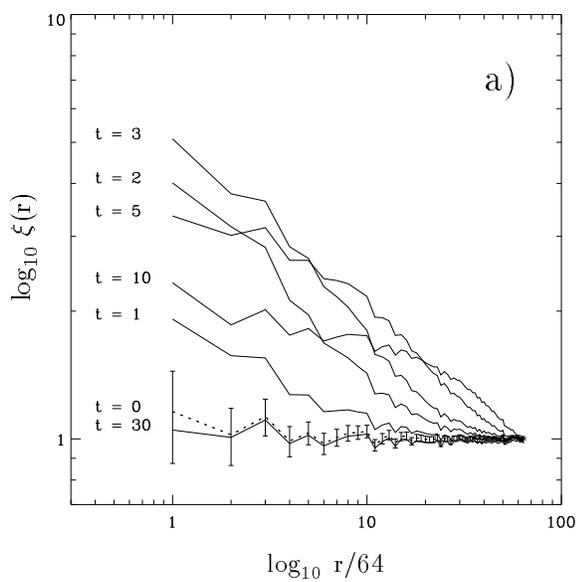
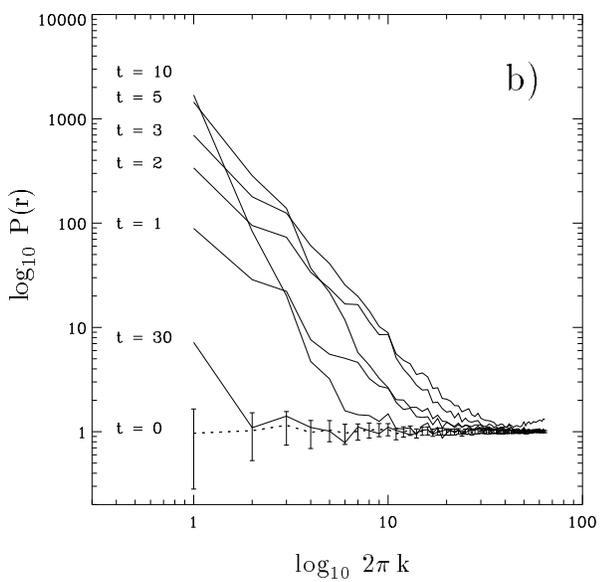
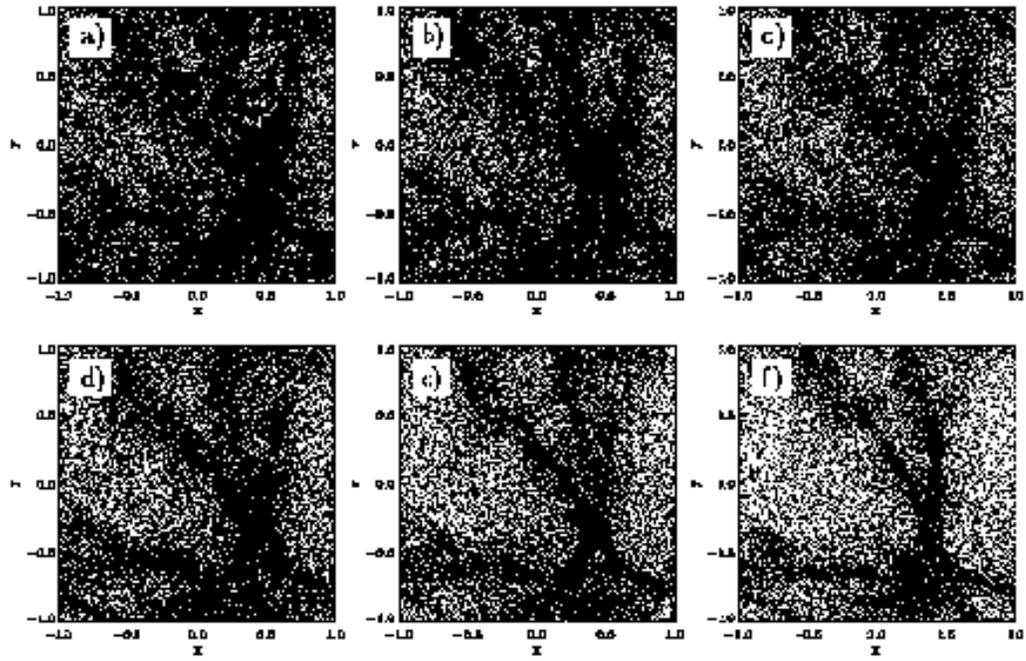
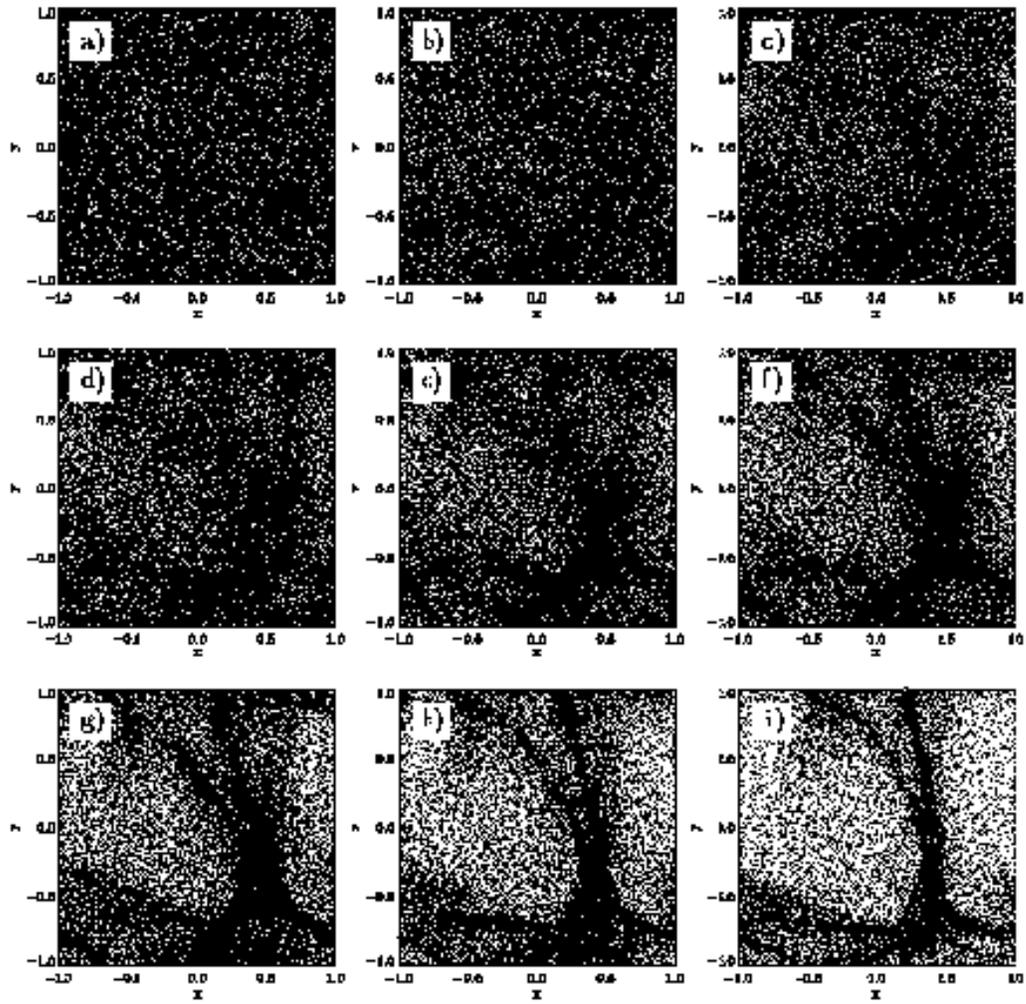
\end{document}